\documentclass{aa}  

\usepackage{graphicx}
\usepackage{txfonts}
\usepackage{makecell}
\usepackage{subfiles}
\usepackage{siunitx}
\usepackage[normalem]{ulem}
\usepackage{threeparttable}

\usepackage{etoolbox}
\patchcmd{\maketitle}{\cleardoublepage}{}{}{}
\patchcmd{\maketitle}{\clearpage}{}{}{}

\newcommand\teff{\ensuremath{T_{\textup{eff}}}\xspace}

\newcommand\muHz{\ensuremath{\mu \mathrm{Hz}}\xspace}
\newcommand\numax{\ensuremath{\nu_{\textup{max}}}\xspace}
\newcommand\dnu{\ensuremath{\Delta\nu}\xspace}

\newcommand\FeH{\ensuremath{[\mathrm{Fe}/\mathrm{H}]}\xspace}

\newcommand\ech{{\'e}chelle\xspace}

\newcommand\deltapi{\ensuremath{\Delta \Pi}\xspace}
\newcommand\solarmass{\ensuremath{M_\odot}\xspace}
\newcommand\solarradius{\ensuremath{R_\odot}\xspace}

\newcommand\rogue{Rogue\xspace}
\newcommand\hennes{Hennes\xspace}
\newcommand\kicrogue{KIC\,7693833\xspace}
\newcommand\kichennes{KIC\,4671239\xspace}

\newcommand\gaia{\textit{Gaia}\xspace}
\newcommand\kepler{\textit{Kepler}\xspace}

\newcommand\basta{\textsc{basta}\xspace}
\newcommand\famed{\textsc{FAMED}\xspace}
\newcommand\diamonds{\textsc{Diamonds}\xspace}

\begin{document} 
    \title{Pushing the boundaries of asteroseismic individual frequency modelling: unveiling two evolved very low-metallicity red giants}

   \author{J. R. Larsen\inst{1}\fnmsep\thanks{E-mail: jensrl@phys.au.dk}
            \and 
            J. L. Rørsted \inst{1,2}
            \and
            V. Aguirre Børsen-Koch \inst{3}
            \and
            M. S. Lundkvist\inst{1}
            \and
            J. Christensen-Dalsgaard\inst{1}
            \and
            M. L. Winther \inst{1}
            \and
            A. Stokholm \inst{4,1}
            \and
            Y. Li \inst{5}
            \and 
            D. Slumstrup \inst{6,7,8}
            \and 
            H. Kjeldsen\inst{1,9}
            \and
            E. Corsaro \inst{10}
            \and
            O. Benomar \inst{11,12}
            \and
            S. Dhanpal \inst{13}
            \and
            A. Weiss \inst{14}
            \and
            B. Mosser \inst{15}
            \and 
            S. Hekker \inst{16,17,1}
            \and
            D. Stello \inst{18, 19, 20, 1}
            \and
            A. J. Korn \inst{21}
            \and
            A. Jendreieck \inst{22}
            \and
            Y. Elsworth \inst{4}
            \and
            R. Handberg\inst{1}
            \and 
            T. Kallinger \inst{23}
            \and 
            C. Jiang \inst{24}
            \and 
            G. Ruchti \inst{25}\fnmsep\thanks{Greg sadly passed away before the completion of this manuscript. We acknowledge his contributions and extend our condolences to his family.}
        }
        
   \institute{Stellar Astrophysics Centre (SAC), Department of Physics and Astronomy, Aarhus University,
              Ny Munkegade 120, 8000 Aarhus C, Denmark 
              \and
              Aarhus Astronomy Data Centre (AADC), Department of Physics and Astronomy, Aarhus University,
              Ny Munkegade 120, 8000 Aarhus C, Denmark 
              \and 
              DARK, Niels Bohr Institute, University of Copenhagen, Jagtvej 128, 18, 2200, Copenhagen, Denmark 
              \and
              School of Physics and Astronomy, University of Birmingham, Edgbaston B15 2TT, UK 
              \and 
              Institute for Astronomy, University of Hawai´i, 2680 Woodlawn Drive, Honolulu, HI 96822, USA 
              \and
              GRANTECAN, Cuesta de San José s/n, E-38712, Breña Baja, La Palma, ES
              \and
              Instituto de Astrofísica de Canarias, E-38205 La Laguna, Tenerife, ES
              \and
              European Southern Observatory, Alonso de Cordova 3107, Vitacura, Chile 
              \and
              Aarhus Space Centre (SpaCe), Department of Physics and Astronomy, Aarhus University, Ny Munkegade 120, 8000 Aarhus C, Denmark 
              \and
              INAF – Osservatorio Astrofisico di Catania, Via S. Sofia, 78, 95123 Catania, Italy
              \and
              Department of Astronomical Science, School of Physical Sciences, SOKENDAI, 2-21-1 Osawa, Mitaka, Tokyo, 181-8588, Japan
              \and
              Solar Science Observatory, National Astronomical Observatory of Japan, 2-21-1 Osawa, Mitaka, Tokyo, 181-8588, Japan
              \and
              Scientific Machine Learning group, Rutherford Appleton Laboratory, Science and Technology Facilities Council, Harwell Campus, Didcot OX11 0QX, UK
              \and 
              Max-Planck-Institute for Astrophysics, Karl-Schwarzschild-Str. 1, 85748 Garching, Germany
              \and
              LIRA, Observatoire de Paris, Universit\'e PSL, CNRS, Sorbonne Universit\'e, Universit\'e Paris-Cit\'e, 92195 Meudon, France
              \and
              Heidelberger Institut für Theoretische Studien, Schloss-Wolfsbrunnenweg 35, 69118 Heidelberg, Germany
              \and
              Center for Astronomy (ZAH/LSW), Heidelberg University, Königstuhl 12, 69117 Heidelberg, Germany
              \and
              School of Physics, University of New South Wales, NSW 2052, Australia
              \and 
              Sydney Institute for Astronomy (SIfA), School of Physics, University of Sydney, NSW 2006 Australia
              \and
              ARC Centre of Excellence for All Sky Astrophysics in 3 Dimensions (ASTRO 3D), Australia
              \and
              Division of Astronomy and Space Physics, Dept. of Physics and Astronomy, Uppsala University, Box 516, 75120 Uppsala, Sweden
              \and 
              Independent Researcher, Münich, Germany -- maxie.jendreieck@gmail.com
              \and
              Institut f\"ur Astrophysik, Universit\"at Wien, T\"urkenschanzstrasse 17, 1180 Vienna, Austria
              \and 
              Max-Planck-Institut f{\"u}r Sonnensystemforschung, Justus-von-Liebig-Weg 3, 37077 G{\"o}ttingen, Germany
              \and 
              Lund Observatory, Department of Astronomy and Theoretical Physics, Box 43, SE-221 00 Lund, Sweden
            }
   \date{Received 21 December 2024; Accepted 28 March 2025}

  \abstract
   {Metal-poor stars play a crucial role in understanding the nature and evolution of the first stellar generation in the Galaxy. Previously, asteroseismic characterisation of red-giant stars has relied on constraints from the global asteroseismic parameters and not the full spectrum of individual oscillation modes. Using the latter, we present for the first time the characterisation of two evolved very metal-poor stars including the detail-rich mixed-mode patterns.
   }
   {We will demonstrate that incorporating individual frequencies into grid-based modelling of red-giant stars enhances its precision, enabling detailed studies of these ancient stars and allowing us to infer the stellar properties of two very metal-poor \FeH~${\sim}{-}2.5$ dex {\it Kepler} stars: KIC 4671239 and KIC 7693833.
   }
   {Recent developments in both observational and theoretical asteroseismology have allowed for detailed studies of the complex oscillation pattern of evolved giants. In this work, we employ \textit{Kepler} time series and surface properties from high-resolution spectroscopic data within a grid-based modelling approach to asteroseismically characterise KIC 4671239 and KIC 7693833 using the BAyesian STellar Algorithm, \basta.}
   {Both stars show agreement between constraints from seismic and classical observables; an overlap unrecoverable when purely considering the global asteroseismic parameters. \kichennes and \kicrogue were determined to have masses of $0.78^{+0.04}_{-0.03}$ and $0.83^{+0.03}_{-0.01} \ M_{\odot}$ with ages of $12.1^{+1.6}_{-1.5}$ and $10.3^{+0.6}_{-1.4}$ Gyr, respectively. Particularly, for \kichennes the rich spectrum of model frequencies closely matches the observed. 
   }
  {A discrepancy between the observed and modelled \numax of ${\sim}10\%$ was found, indicating a metallicity dependence of the \numax scaling relation. For     metal-poor populations, this results in overestimations of the stellar masses and wrongful age inferences. Utilising the full spectra of individual oscillation modes lets us circumvent the dependence on the asteroseismic scaling relations through direct constraints on the stars themselves. This allows us to push the boundaries of state-of-the-art detailed modelling of evolved stars at metallicities far different from solar.}

   \titlerunning{Pushing the boundaries of asteroseismic individual frequency modelling}  
   \maketitle

\clearpage
\section{Introduction}\label{sec:Intro}
The available archive of observed red-giant stars has been expanded significantly by the \kepler space mission \citep{Borucki10}, allowing for many detailed studies to be carried out in its wake. These studies focus on the classification and understanding of red giants by the inferences possible from their oscillatory characteristics (e.g., \citealt{Jiang14}; \citealt{Mosser18}; \citealt{Lindsay22}). The nature of these stellar pulsations is studied by the field of asteroseismology; the application of which has been revolutionised by the wealth of available time series data from \kepler. They allow for so-called seismic determinations of the fundamental stellar parameters such as mass, radius and surface gravity $g$ -- but importantly also the possibility of an accurate inference on the stellar age (see e.g. \citealt{Soderblom15}).

Until now, these seismic determinations of red giants have relied on extensive application of the asteroseismic scaling relations (e.g., \citealt{Pinsonneault14, Handberg17, Brogaard22, Pinsonneault24}), which relate observed asteroseismic quantities to the stellar mass and radii\citep{Kjeldsen95}. The scaling relations are based on two global asteroseismic properties: the frequency of maximum oscillatory power \numax, which is related to the surface gravity \citep{Brown91,Belkacem11}, and \dnu, the average large frequency separation between consecutive modes of identical spherical degree $\ell$, owing a dependence to the mean density of the star \citep{Tassoul80}. The application of the asteroseismic scaling relations, known as the so-called \emph{direct method} in the literature (for a review see \citealt{Hekker20}), entails that the inferences drawn carry the crucial assumption that we can scale the fundamental parameters of a star according to well-known solar values. In the asymptotic regime these relations are \citep{Chaplin13}: 
\begin{align}
    \dnu &\simeq \left(\frac{M}{M_{\odot}} \right)^{1/2} \left( \frac{R}{R_{\odot}}\right)^{-3/2} \Delta\nu_{\odot} \label{eq:DnuScal}\\
    \numax &\simeq \left( \frac{M}{M_{\odot}}\right)\left(\frac{R}{R_{\odot}}\right)^{-2} \left( \frac{\teff}{T_{\mathrm{eff},\odot}}\right)^{-1/2}\nu_{\mathrm{max},\odot} \propto \frac{g}{\sqrt{T_\mathrm{eff}}} . \label{eq:NumaxScal}
\end{align}
The benefit of applying Eqs.~\ref{eq:DnuScal} and \ref{eq:NumaxScal} is a model-independent determination of the stellar mass and radius if a measurement of the effective temperature \teff is available. They have therefore been applied widely to \kepler red giants near solar metallicity. However, \kepler also discovered metal-poor giants. The application of the direct method for such stars was studied by \citet{Epstein14} for a sample of nine metal-poor giants. Employing the scaling relations with effective temperatures from the Apache Point Observatory Galactic Evolution Experiment (APOGEE; \citealt{Accetta22}), the authors found a systematic overestimation in the recovered masses compared to the expectation from e.g. fits to the colour-magnitude diagrams of globular clusters. The reliability of the scaling relations for red giants at compositions different than solar has therefore become questionable. 

An alternative to the direct method is grid-based modelling, where the information on the composition can be incorporated by fitting the observed atmospheric and asteroseismic values to grids of stellar models. Thereby, theoretical values for the global asteroseismic parameters can be predicted from the models and compared to the observed counterparts \citep{Stello09, Aguirre15, Aguirre22, Puls22, Huber24}. Matching the observed values to the grids means that grid-based modelling has a critical dependence on our ability to infer accurate and reliable theoretical values of the atmospheric and global asteroseismic parameters for the stellar models. In case of \dnu one can either employ Eq.~\ref{eq:DnuScal} or obtain it from theoretically computed individual frequencies for the models. However, for \numax the only option available for a theoretical estimate is to employ Eq.~\ref{eq:NumaxScal}, where one may include several correction terms as a function of metallicity (e.g. \citealt{Viani17}). The inability of estimating \numax directly from the models stems from the stochastically excited nature of the oscillations themselves, where the non-adiabatic interplay between excitation (energy supplied from turbulent convection) and damping rate would be required to evaluate the amplitude of the oscillations \citep{Houdek06}. 

In this work, for the first time within the field, the grid-based modelling of two evolved ($\numax \lesssim 100$ \muHz as classified by \citealt{Huber24}) very low-metallicity  red-giant stars will rely on fitting the complete range of $\ell={0,1,2}$ observed individual frequencies, with the inclusion of the mixed modes. In red giants, any non-radial ($\ell\neq0$) individual frequencies will display mixed-mode characteristics. This occurs due to the possibility of a coupling between the interior g- and exterior p-mode properties, when the respective oscillation cavities exist at identical frequency ranges near \numax \citep{Jiang14}. The strength of the coupling is measured through the coupling constant $q$.
The identification of such mixed modes means observing pressure modes that have been perturbed by the interior gravity-wave behaviour, thus carrying diagnostic information valuable for constraining the interior of the red-giant star. Modelling red giants through fitting directly to the individual frequencies will thereby circumvent the aforementioned dependence on the scaling relations; ideally providing the resulting stellar models whose interior profiles best match the constraints on the stellar interior supplied by the oscillation frequencies. This expansion was made computationally efficient by the recent developments of \citet{Larsen24}, aiding the availability of the theoretical individual frequencies for red-giant stellar models in the context of large-scale grids. 

Previously, \citet{Ball18} employed a similar technique to fit the individual frequencies of three RGB stars, albeit with the goal of studying the surface effects for red giants. However, in their scheme they did not attempt to utilise the mixed modes of the models. This means that they had to match the observed frequencies to the single most p-mode-like model frequency (acoustic resonance) within each mode order. Hence, they lost the information on the deep stellar interior carried by the mixed modes and the asymptotic dipole mode period spacing $\Delta\Pi_1$. \citet{Campante23} performed detailed modelling of three evolved RGB stars near solar metallicity, having access to the complete dipole spectrum in the models but limited to the p-mode resonances for the quadrupoles, aiming to test the impact of different sets of observable frequencies as constraints. Recently, \citet{Huber24} presented detailed modelling fitting the individual frequencies of a less evolved ($\numax \sim 200$ \muHz) metal-poor giant KIC 8144907, proving that stellar modelling of such stars is possible and, moreover, asteroseismically characterising the most metal-poor giant to date. In this work, we further extend the considerations and the approach of grid-based individual frequency modelling for similarly metal-poor, yet notably more evolved red giants with $\numax \lesssim 100$ \muHz. 

This project is the culmination of many years of efforts begun in the early 2010's, firstly by \citet{Jendreieck12} and later led by V. Aguirre Børsen-Koch (formerly Silva Aguirre) and completed in the present work. The initial efforts focused on proper characterisation of two, to become notoriously vexing, very low-metallicity red-giant stars: KIC 4671239 (dubbed "Hennes") and KIC 7693833 (dubbed "Rogue"). These stars resisted proper and coherent characterisation in modelling efforts, and a thorough investigation effort was therefore undertaken to obtain atmospheric and asteroseismic determinations for both, before attempting to perform grid-based modelling. The project never reached completion and remained unpublished, primarily due to the lack of coherence between the fitted results and observational constraints. Additionally, both stars were recovered with solar-like ages, which conflicts with expectations from galactic archaeology that very low-metallicity (possibly halo) stars should be old. Certain elements from the prior efforts by V. Silva Aguirre et al. (unpublished) have been adopted into our present analysis.

KIC 4671239, henceforth Hennes, is one of the most metal-poor ($\FeH\sim -2.6$ dex) stars found in the \emph{Kepler} field. As an individual star, Hennes has been studied several times in different contexts, first classified in the SAGA survey by \citet{Casagrande14} using Strömgren photometry. \citet{Mosser17} investigated the coupling factor $q$ of mixed modes for red giants, and noted that specifically for Hennes it was found to be atypically high, resulting in a complex mixed-mode pattern. Furthermore, they determined a dipole spacing $\Delta\Pi_1 \approx 66.6$~s -- which characterises Hennes as a red-giant branch (RGB) star according to the work of \citet{Bedding11} -- that they note is much smaller than for other RGB stars of similar \dnu. Hennes may then resemble stars identified by \citet{Deheuvels22} as resulting from mass transfer, which would result in the inability of canonical stellar models describing the star. In this work, however, we find that the very low metallicity is the cause of the situation (see Sect.~\ref{subsec:InteriorMixing}). An attempt to characterise Hennes using grid-based modelling with global asteroseismology, along with other giants, was performed by \citet{Puls22}. They recovered a solar-like age for Hennes, and speculated the possibility for Hennes to be a blue straggler or potentially having experienced a merger event in its past. Recently, the ensemble study by \citet{Kuszlewicz23} also recovered a high coupling factor $q\sim0.3$ for Hennes, proposing the possibility of high coupling values being an indicator of metal-poor giants.

\begin{figure}[t]
    \resizebox{\hsize}{!}{\includegraphics[width=\linewidth]{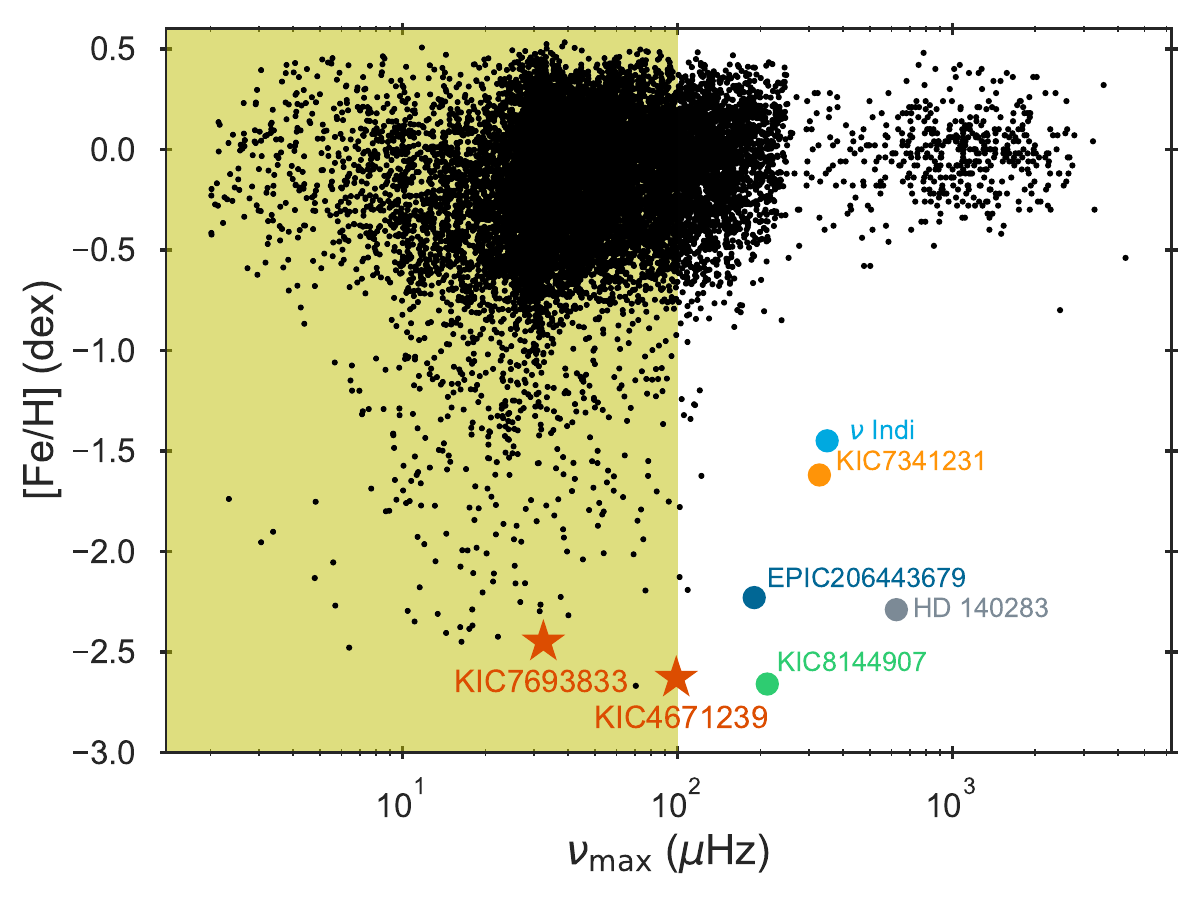}}
    \centering
    \caption{Metallicity \FeH plotted against the frequency of maximum oscillatory power \numax for known oscillating stars, adapted from the left panel of Fig.~2 from \citet{Huber24}. The shaded area represents the region of evolved RGB stars, with a boundary set at $\numax \lesssim 100$ \muHz. Noteworthy low-metallicity sub-giant branch (SGB) and RGB stars for which asteroseismic modelling has been performed are highlighted. Note that EPIC206443676 was modelled using the global asteroseismic parameters \citep{Deheuvels12}, not the individual frequencies as in all other displayed cases. The red stars indicate the two giants considered in the present work.}
    \label{fig:Overview}
\end{figure}

KIC 7683833, hereafter Rogue, is a similarly metal-poor ($\FeH\sim -2.4$ dex) giant initially identified by \citet{Thygesen12} and preliminarily studied by \citet{Jendreieck12}. Rogue has not been studied as an individual star since, but it has been classified by ensemble studies as an RGB star (\citealt{Vrard16}; \citealt{Elsworth16}; \citealt{Mathur17}; \citealt{Yu18}). The most recent ensemble work of \citet{Pinsonneault24} (the APOKASC 3 catalogue), using global asteroseismic parameters and the scaling relations, recovered a mass of $\sim1.05$ M$_\odot$ and age of $\sim4.85$ Gyr for Rogue. These results of a solar-like mass and age, similarly to the prior efforts, are difficult to reconcile with astrophysical expectations.

Fig.~\ref{fig:Overview} is adapted from \citet{Huber24}\footnote{The literature values that form the black points in the plot are from \citet{Pinsonneault14}, \citet{Serenelli17}, \citet{Matsuno21} and \citet{Schonhut-Stasik24}.} and highlights the evolved nature of Hennes and Rogue alongside the extension to the previous situation that this work contributes. The shaded region represents evolved giants where individual frequency modelling is challenging to perform (see e.g, Fig~2 of \citealt{Larsen24}). The small existing selection of asteroseismically modelled and characterised low- and very low-metallicity stars are seen in the figure, but all lie outside of the evolved region. Hennes and Rogue push into this region, extending the boundaries of individual frequency modelling, while also being the first very low-metallicity evolved red giants to be asteroseismically characterised. 

The paper is structured as follows: in Sect.~\ref{sec:ClassicParams} we outline the spectroscopic and photometric observations of both stars, before continuing to the asteroseismic observations and reduction in Sect.~\ref{sec:Asteroseismic}. The details of the grid-based modelling is presented in Sect~\ref{sec:GridModelling}, with details on the stellar evolution and oscillation codes used. Furthermore, we briefly introduce the BAyesian STellar Algorithm (BASTA; \citet{Aguirre22}) used within this work to perform the fitting. Section~\ref{sec:Results} presents the results obtained for Hennes and Rogue, before discussing their implications and further context in Sect.~\ref{sec:Discussion}. Lastly, we conclude in Sect.~\ref{sec:Conclusion}.

\section{Classical parameters}\label{sec:ClassicParams}

\begin{table*}[]
\renewcommand{\arraystretch}{1.1}
\centering
\caption{Compiled spectroscopic determinations for Hennes and Rogue from various sources. Note that the value for \teff and \FeH from \citet{Yu18} for Hennes are adopted from \citet{Casagrande14}. For Rogue, \citet{Yu18} adopted the corresponding values from \citet{Mathur17}, who in turn adopted them from \citet{Thygesen12}. The estimates from APOGEE SDSS DR17 \citep{Accetta22} are also given for Rogue, estimated from a single spectrum (visits by APOGEE). Note that the estimate for $[\alpha/\mathrm{Fe}]$ for this entry is given as the median of the individual Mg and Si abundances, with the uncertainties added in quadrature. The values adopted for each star in the modelling of this work are highlighted in bold.}
\begin{threeparttable}
\begin{tabular}{l|llll}
\hline
\multicolumn{1}{c|}{Source}  & \multicolumn{1}{c}{\teff [K] }  & \multicolumn{1}{c}{$\log(g)$}  & \multicolumn{1}{c}{\FeH [dex]}  & \multicolumn{1}{c}{$[\alpha/\mathrm{Fe}]$ [dex]} \\ \hline 
\multicolumn{1}{c|}{}  &  \multicolumn{4}{c}{\textbf{Hennes -- KIC 4671239}} \\ \hline
\citet{Casagrande14}  & 5224 $\pm$ 104  &  2.92 $\pm$ 0.004   & -2.44 $\pm$ 0.17  &  --               \\
\citet{Yu18}          & 5224 $\pm$ 100  &  2.922 $\pm$ 0.006  & -2.44 $\pm$ 0.30  &  --               \\
\textbf{\citet{Puls22}}        & \textbf{5295} $\mathbf{\pm}$ \textbf{145}  &  \textbf{2.929} $\mathbf{\pm}$ \textbf{0.15}   & \textbf{-2.63} $\mathbf{\pm}$ \textbf{0.20}  &  \textbf{0.125} $\mathbf{\pm}$ \textbf{0.25} \\
Sect.~\ref{subsec:Atmospheric}  & 5080 $\pm$ 220  & 2.93 $\pm$ 0.10     & -2.71 $\pm$ 0.11  & 0.12 $\pm$ 0.16     \\
\hline
\multicolumn{1}{c|}{}  &  \multicolumn{4}{c}{\textbf{Rogue -- KIC 7693833}} \\ \hline
\citet{Thygesen12}    & 4880 $\pm$ 100  &  2.46 $\pm$ 0.01    & -2.23 $\pm$ 0.15  &  --               \\
\citet{Yu18}          & 4880 $\pm$ 97   &  2.413 $\pm$ 0.008  & -2.23 $\pm$ 0.15  &  --               \\
APOGEE     & 5026 $\pm$ 17.2 &  2.41 $\pm$ 0.06    & -2.33 $\pm$ 0.01  &  0.32 $\pm$ 0.05 \tnote{*}  \\
\textbf{Sect.~\ref{subsec:Atmospheric}} & \textbf{4840} $\mathbf{\pm}$ \textbf{150}  &  \textbf{2.43} $\mathbf{\pm}$ \textbf{0.10}    & \textbf{-2.45} $\mathbf{\pm}$ \textbf{0.05}  & \textbf{0.15} $\mathbf{\pm}$ \textbf{0.12} \\
\hline
\end{tabular}
\begin{tablenotes}\footnotesize
\item[*] Note that the uncertainties reported by the ASPCAP pipeline in APOGEE are purely internal and thus do not account for systematics, resulting in unrealistically low uncertainties (Holtzman et al in prep). 
\end{tablenotes}
\end{threeparttable}
\label{tab:SpecComp}
\end{table*}

The various sources for the atmospheric and photometric parameters relevant for the modelling of Hennes and Rogue are outlined in the following. As mentioned in Sect. \ref{sec:Intro} there are estimations available for both stars from independent determinations. All literature values for the spectroscopic and photometric parameters of interest are listed in Table \ref{tab:SpecComp}. The nuances and procedures for the estimates originating from this work will be presented below.

\subsection{Atmospheric observations}\label{subsec:Atmospheric}
The atmospheric analysis is performed on observations taken at the Nordic Optical Telescope (NOT) in 2012/13. Both stars were observed with the FIES spectrograph, which is a fixed-setting spectrograph covering a spectral range of 370-730 nm without gaps \citep{FIES}. The observations were carried out with the medium-resolution setting, resulting in a resolving power of $R=46000$. Hennes was observed on a single night in October 2013, obtaining four spectra of an hour each. Rogue was observed on two nights in August 2012, obtaining two spectra of an hour each. Additionally, for Rogue we obtained new high-resolution ($R=67000$) FIES spectra with a combined exposure time of $\sim$4.8 hours with the NOT in May 2024 in an effort to better characterise the star. The results of this work for Rogue in Table \ref{tab:SpecComp} are obtained from these newer spectra.

The spectroscopic analysis follows the method outlined in \citet{slumstrup2019} by using the asteroseismically determined surface gravity to constrain a classical equivalent width analysis of Fe lines. The effective temperature is varied until excitation equilibrium is reached, i.e. no trend is observed between the abundance and the excitation potential of the Fe lines. Likewise, the microturbulence is determined by removing the trend between the abundance and the strength of the Fe lines. Uncertainties were calculated by varying each parameter until at least a 3$\sigma$ uncertainty is produced on the slope of [Fe/H] vs. excitation potential. Agreement between abundances of FeI and FeII lines to within one standard deviation was assured throughout the analysis.
The equivalent widths were measured with DAOSPEC \citep{stetson2008} and the abundances were calculated with the auxiliary program \textit{Abundance} with SPECTRUM in LTE \citep{Gray1994}. The stellar atmosphere models used are ATLAS9 \citep{castelli2004} with solar abundances from \citet{grevesse1998}. The stars analysed in the present work are significantly more metal-poor than those of \citet{slumstrup2019} and therefore it has been necessary to implement NLTE departure-coefficients taken from the INSPECT database version 1.0\footnote{Available at \url{www.inspect-stars.com}} \citep{Bergemann2012, Lind2012}. The alpha enhancement is determined from the Mg abundance with NLTE departure-coefficients.

The metallicities \FeH derived in this work are on average lower than the alternative measurements. However, this is in coherence with expectations as the derived temperatures are also lower. In this parameter space, a lower temperature will provide stronger spectral lines, effectively cancelling out the effect of a lower metallicity. Conversely, in the other estimates the temperatures and metallicities are higher, once again counteracting each other and giving coherent results. The determining factor then becomes the microturbulence as a fitting parameter, which becomes difficult to implement in this very low-metallicity regime for a 1D NLTE approach. The variation seen in $[\alpha/\mathrm{Fe}]$ between our results and APOGEE for Rogue likely stems from the different reduction methods, namely equivalent widths and synthetic spectra, respectively. Spectroscopic reduction in the very low-metallicity regime is difficult for both our analysis and the ASPCAP pipeline, as both are optimised for higher metallicities. The comparably higher resolution of the NOT spectra allows for a better determination of the individual spectral line strengths, which is advantageous for this metallicity regime, where so few lines are present.

In the case of Hennes, the choice of spectroscopy required additional considerations. The choice was between our own results from the NOT observations of 2013 and the results of \citet{Puls22} from the HIRES spectrograph \citep{Vogt94}. The two datasets are consistent within the uncertainties, yet when comparing to purely spectroscopic test-fits to stellar atmospheric models, a self-consistent overlap between all parameters and the models only occur for the HIRES spectroscopy of \citet{Puls22}. Furthermore, HIRES is mounted on the larger W. M. Keck telescope and the observations have a higher signal-to-noise ratio than the NOT counterparts for Hennes. We therefore choose to employ the spectroscopic determinations of \citet{Puls22} for the modelling in Sects.~\ref{sec:GridModelling} and \ref{sec:Results}. The estimate of the alpha-enhancement $[\alpha/\mathrm{Fe}]$ is calculated as the mean of the two alpha-element abundances from magnesium and calcium, with the uncertainty found by adding the individual element uncertainties in quadrature. 

For Rogue we choose to use the spectroscopic determination from our present work. They are consistent with the other estimates within the uncertainties. As argued above, the differences seen in Table~\ref{tab:SpecComp} are likely caused by the difficult parameter space of the star for 1D NLTE approaches, and indicates that both stars would likely benefit from a full 3D NLTE analysis in the future.

\begin{figure*}[t]
    \includegraphics[width=17cm]{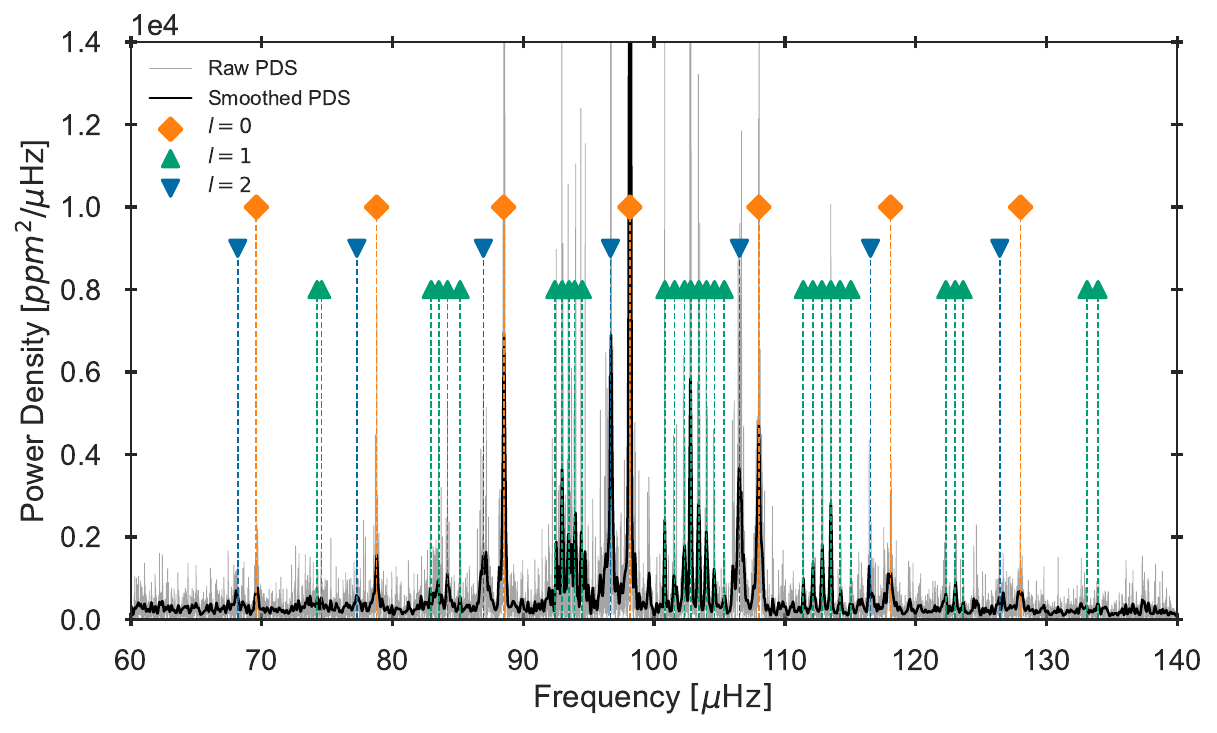}
    \centering
    \caption{Power density spectrum of Hennes with the consolidated list of individual frequencies from Table~\ref{tab:FreqAsteroHennes} overplotted. Both the raw spectrum (shaded gray) and a Gaussian smoothed version (black) are seen. The orange diamonds, green upwards and blue downwards triangles depict the $\ell=0$, $\ell=1$ and $\ell=2$ modes, respectively. }
    \label{fig:HennesPDS}
\end{figure*}

\subsection{Photometric and astrometric observations}
In order to obtain constraints on the stellar luminosities (see \citealt{Aguirre22} for details), we need a measure for the parallax and observed magnitudes. In \gaia DR3 \citep{Gaia16,GaiaDR3}, the 5-parameter astrometric solution was measured for both Hennes and Rogue as well as the \gaia $G$, $G_{\textup{BP}}$, and $G_{\textup{RP}}$ magnitudes, hence we can add these to the inference of stellar properties. Hennes has a magnitude of 13.6 in the G band, while Rogue has a magnitude of 11.7. The uncertainties of the \gaia magnitudes are internal and extremely small. A floor of $\sigma=0.01$ has therefore been set for all magnitudes retrieved. The \gaia parallaxes were also recovered and have been corrected for their known zero-point error according to the description of \citet{Lindegren21}. This results in a corrected parallax of $0.471$ and $0.659$ mas for Hennes and Rogue, respectively.

\section{Asteroseismic observations and data reduction}\label{sec:Asteroseismic}
Hennes and Rogue were observed in the nominal \kepler mission. The time series were recovered from the KASOC database\footnote{\url{https://kasoc.phys.au.dk}} after reduction by the pipeline \citep{Handberg14}. Subsequently, they were reduced to form the power density spectra (PDS) of both stars following the prescription by \citet{Handberg11}. Several independent analyses of the PDS were performed, providing primarily the observed individual frequencies for comparison and cross-referencing. Depending on the analysis employed, estimates of the global asteroseismic parameters were also returned. The nuances of the independent determinations and their results are described in Appendix~\ref{app:A}, where Table~\ref{tab:ColabGlobParams} summarises the global asteroseismic parameter estimates obtained. Subsequently, we briefly describe the procedure for obtaining the global asteroseismic parameters in this work.

The value of \numax was found by assuming that the power excess represents a Gaussian envelope \citep{Bedding14_obs_perspectives}, where the vertex defines \numax. Repeatedly applying a Gaussian smoothing with a large standard deviation to the PDS allows for the production of the envelope and the subsequent retrieval of \numax for each of the stars. The uncertainty was estimated by varying the length of the time series that formed the PDS and changing the width of the Gaussian, meanwhile recording the variations in the derived value for \numax. These estimates agree with those outlined in Appendix~\ref{app:A}.

For \dnu we apply the method of autocorrelation to recover the periodicity stemming from the even spacing in frequency of the p-modes in the PDS. From the autocorrelation function\footnote{Calculated with the \textsc{StatsModels} module ACF \url{https://www.statsmodels.org/stable/generated/statsmodels.tsa.stattools.acf.html}.}, the peaks corresponding to multiples of \dnu are extracted and a linear correlation through the origin is fitted. The gradient of this linear fit provides \dnu with the uncertainty estimated from the associated covariance matrix. Once more, the values of \dnu obtained corresponds well with the other independent determinations. 

The dipole period spacing $\Delta\Pi_1$ was determined by employing a de-convolution of the gravity-mode pattern from the mixed-mode pattern \citep{Mosser15}, which takes into account the signature of a buoyancy glitch \citep{Cunha15, Lindsay22, Cunha24}. The obtained uncertainty accounts for aliasing effects as described in Appendix~A of \citet{Vrard16}. This framework works for Hennes, and returns a value consistent with the independent determinations. Due to the more evolved nature of Rogue, the proximity between the observed dipole modes exceeds the frequency resolution of the PDS. This, in combination with a weak coupling due to a small coupling factor $q$ (see Sect.~\ref{subsec:AsteroRogue}), makes the determination of the dipolar period spacing $\Delta\Pi_1$ very uncertain.

Subsequently, we summarise the obtained parameters for each star and the inferences possible, before presenting the observed individual frequencies.

\subsection{Hennes - KIC 4671239}\label{subsec:AsteroHennes}
\begin{figure}[t]
    \resizebox{\hsize}{!}{\includegraphics[width=\linewidth]{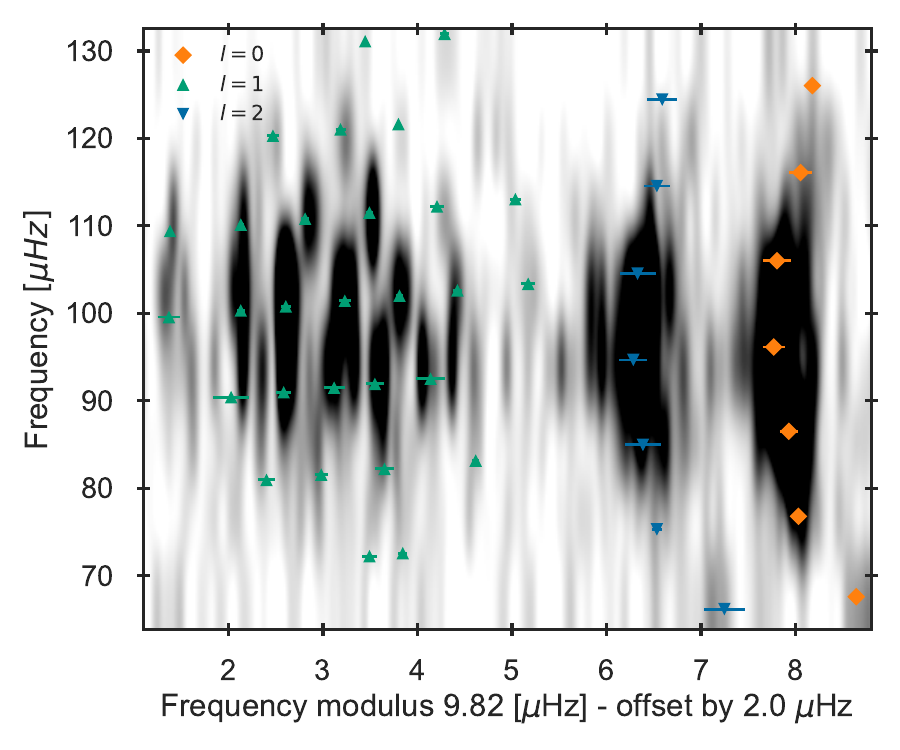}}
    \centering
    \caption{Observed power \ech diagram for Hennes with the consolidated list of individual frequencies from Table~\ref{tab:FreqAsteroHennes} overplotted. The orange diamonds, green upwards and blue downwards triangles depict the $\ell=0$, $\ell=1$ and $\ell=2$ modes, respectively. An offset has been applied to the x-axis for improved visualisation to avoid wrapping of the radial ridge}.
    \label{fig:HennesEchelle}
\end{figure}

\begin{table}[h!]
\centering
\caption{Global asteroseismic parameters for Hennes (KIC 4671239) and Rogue (KIC 7693833).}
\begin{tabular}{lcc
                }
\hline
\multicolumn{1}{l}{\thead{Global parameter}} & \multicolumn{1}{c}{\thead{Hennes}} & \multicolumn{1}{c}{\thead{Rogue}} \\ \hline
\numax [\muHz]            & 98.9 $\pm$ 1.2                & 32.5 $\pm$ 0.2              \\
\dnu [\muHz]            & 9.82 $\pm$ 0.05               & 3.98 $\pm$ 0.01             \\
$\Delta\Pi_1$ [s]         & 66.0 $\pm$ 1.0                & 56.6 $\pm$ 2.0              \\
\hline
\end{tabular}
\label{tab:GlobalAsteroCombined}
\end{table}

\begin{figure*}[t]
    \includegraphics[width=17cm]{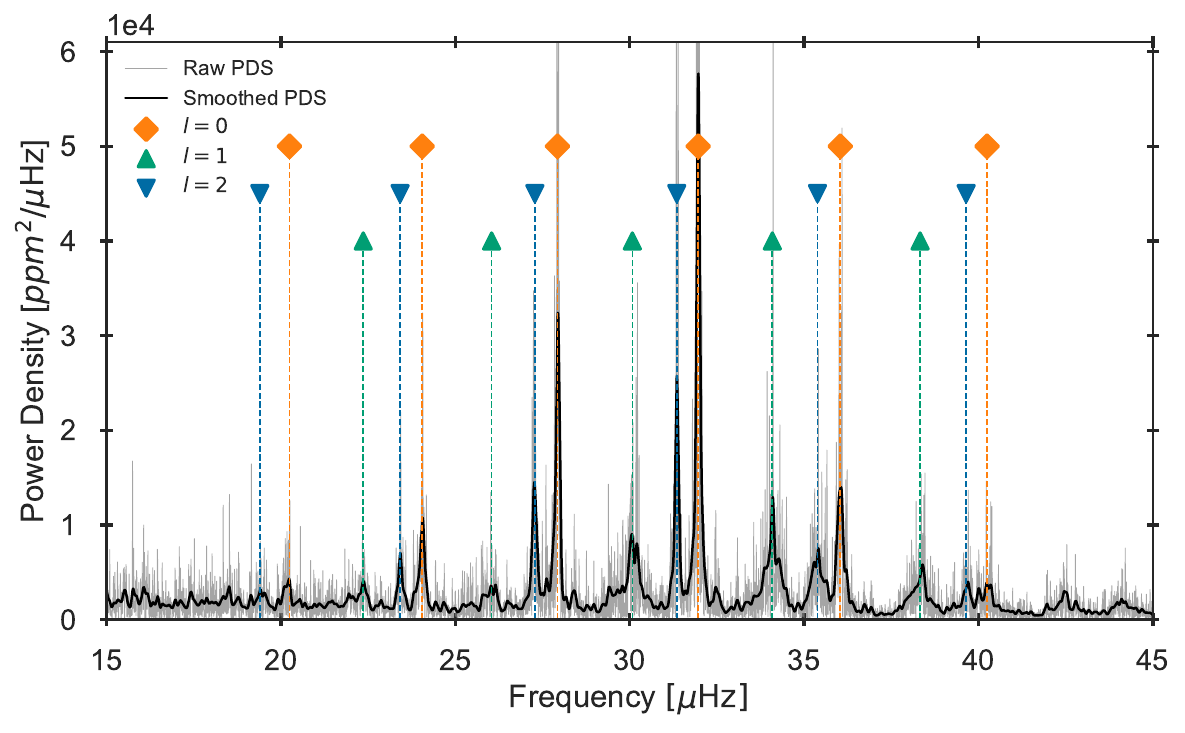}
    \centering
    \caption{Power density spectrum of Rogue with the consolidated list of individual frequencies from Table~\ref{tab:FreqAsteroRogue} overplotted. Both the raw spectrum and a Gaussian smoothed version are shown as the shaded gray and black spectra, respectively. The orange diamonds, green upwards and blue downwards triangles depict the $\ell=0$, $\ell=1$ and $\ell=2$ modes, respectively.}
    \label{fig:RoguePDS}
\end{figure*}

Hennes was observed by \emph{Kepler} from Q1-Q17 in long-cadence mode, with Q6, Q10 and Q14 missing due to the failing of CCD module 3. The resulting timeseries with a length of $\sim$3.5 years produces the PDS seen in Fig.~\ref{fig:HennesPDS}. In this PDS, the dipole mixed modes are clearly separated at higher frequency. This is a result of the unusually high coupling factor of $q=0.26\pm0.03$ found for Hennes, which reflects a small evanescent region in its interior resulting in a strong perturbation to the p-modes by the g-modes (see Sect.~\ref{subsec:InteriorMixing} for a discussion on these aspects). Furthermore, it indicates the less evolved nature of Hennes, where the radiative dampening is not yet strong enough to damp the strongly coupled dipole mixed-modes \citep{Grosjean14}. We note that one of the independent determinations returned an estimated inclination of $57\pm 2$ degrees, which could enforce the possibility of rotational splitting of the modes, tentatively observed for some quadrupole modes in the PDS. Table~\ref{tab:GlobalAsteroCombined} presents the values obtained for the global asteroseismic parameters of Hennes. The star is thus right past the border to being an evolved RGB star (cf. Fig.~\ref{fig:Overview}), confirmed by both the value of \dnu and $\Delta\Pi_1$ \citep{Bedding11}.

The individual frequencies obtained from all independent determinations for Hennes were manually inspected and compared. The frequencies recovered from the FAMED pipeline \citep{Corsaro20} were chosen as they show the most conservative fit to the observed power in the PDS. A cross-match was then performed to only accept frequencies which depicted an overlap within $2\sigma$ to the corresponding mode in another set of determinations. The consolidated list of frequencies are displayed in Table~\ref{tab:FreqAsteroHennes} and show 7 detected acoustic orders, with an additional isolated quadrupole mode at low frequency. This consolidated list was later cross-checked by comparing to the results of the TACO code (Hekker et al. in prep; \citealt{TACOPrelim}, see Appendix~\ref{subapp:TACO}). Figure~\ref{fig:HennesPDS} shows the frequencies overplotted on the PDS, while Fig.~\ref{fig:HennesEchelle} displays the modes in an observed power \ech diagram. An \ech diagram uses the even spacing in frequency of the p-modes to stack the PDS vertically in units of \dnu. In Fig.~\ref{fig:HennesEchelle}, the observed power is plotted in a binned grid to allow for visible inspection of the frequency determinations, which should depict a clear overlap with the power. Figure~\ref{fig:HennesEchelle} clearly illustrates the evolved mixed-mode pattern of Hennes, where a broad selection of dipole modes have been identified across various mode orders. When using the determined frequencies in Table~\ref{tab:FreqAsteroHennes} for the inference in Sect.~\ref{sec:Results}, we apply a correction to the Doppler-shift stemming from the radial velocity $V_r$ of the star as discussed by \citet{Davies14}. This correction factor is calculated for Hennes to be $1+V_r/c = 0.99937$, where $c$ is the speed of light and $V_r$ is $-189.38$ km/s obtained from Gaia DR3 data.

\subsection{Rogue - KIC 7693833}\label{subsec:AsteroRogue}
Rogue was observed through Q0-Q17 in long-cadence mode; the entire nominal mission of \emph{Kepler}, providing a 4-year long timeseries from which the PDS in Fig.~\ref{fig:RoguePDS} is produced. Worth noting is the broadened and substantial power clusters attributed to the dipole modes. In contrast to individual dipole frequency peaks, these clusters arise due to the smaller period spacing and lower coupling strength for this more evolved star; two independent determinations find it to be $q \lesssim 0.09$ (see Table~\ref{tab:ColabGlobParams}; \citealt{Dhanpal22}). The enhances mixing is due to the evolved nature of Rogue, as the oscillation spectrum becomes increasingly dense with evolution \citep{Mosser18,Larsen24}. Furthermore, this results in a slight rightwards shift of the dipole modes from the central position between the $\ell=0$ and $\ell=2$ peaks, as expected from the considerations by \citet{Bedding10} and \citet{Stello14}. 

As shown in Table~\ref{tab:GlobalAsteroCombined}, the values of \dnu and \numax determined for Rogue indicate an RGB star further evolved compared to Hennes, likely situated just prior to the location of the RGB bump \citep{Khan18}. The dipole period spacing was determined to be $\sim56\pm2$ s. However, as introduced earlier, this determination is very uncertain due to the nature of Rogue. The independent determinations found values ranging from $61$ to $105$ seconds, with uncertainties as large as $58$ seconds. Bearing this in mind, the later modelling of Rogue is unable to consider the period spacing as a global parameter for fitting in Appendix~\ref{app:C}.

\begin{figure}[t]
    \resizebox{\hsize}{!}{\includegraphics[width=\linewidth]{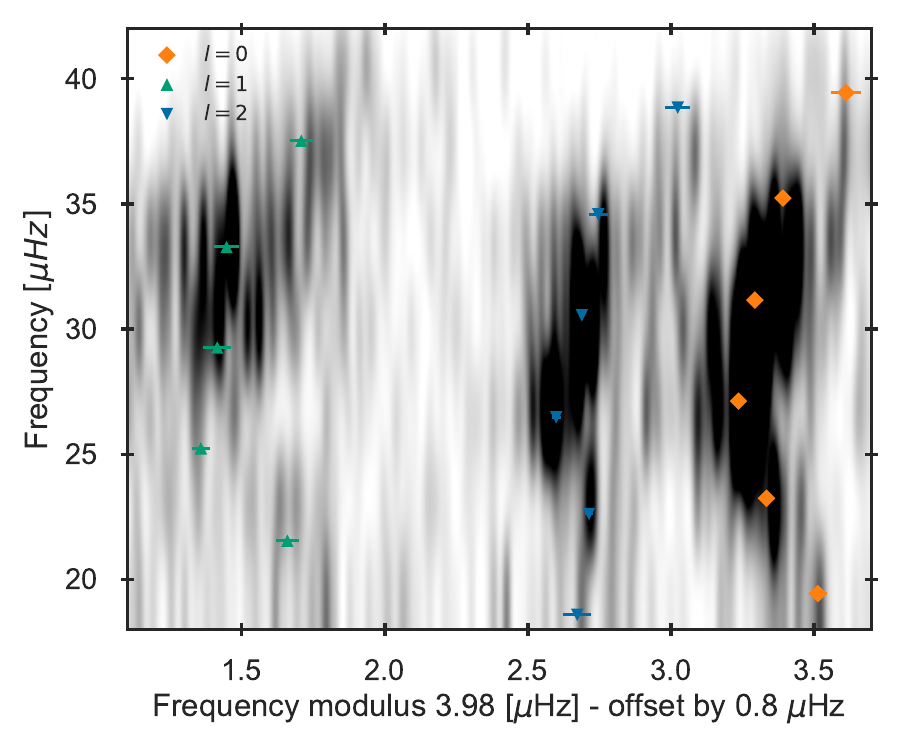}}
    \centering
    \caption{Observed power \ech diagram for Rogue with the consolidated list of individual frequencies from Table~\ref{tab:FreqAsteroRogue} overplotted. The orange diamonds, green upwards and blue downwards triangles depict the $\ell=0$, $\ell=1$ and $\ell=2$ modes, respectively. An offset has been applied to the x-axis for improved visualisation to avoid wrapping of the radial ridge}.
    \label{fig:RogueEchelle}
\end{figure}

The individual frequencies recovered by the approach of \citet{Liyg2020} were selected for Rogue, as the determination of the radial ($\ell=0$) and quadrupole ($\ell=2$) modes overlap with all other independent determinations. Furthermore, the frequency set includes only the single p-dominated dipole mode in each mode order. This is desirable, as the reliability of the dipole mixed-mode determination is questionable for Rogue due to the concerns outlined above regarding the frequency resolution and the low coupling factor. The list of determined frequencies is provided in Table \ref{tab:FreqAsteroRogue} and shows 5 fully detected acoustic orders with two additional isolated modes, one radial and one quadrupole. This consolidated list was also cross-checked against the results of TACO. 

The fitting of $\ell=2$ modes as Lorentzian profiles is under the assumption that they originate from a damped oscillator. This may pose issues due to the presence of mixed modes. Mixed modes can lead the fitting process to converge on a specific mixed mode instead of a pure p-mode. This misidentification can result in a smaller line width and an unrealistically low frequency uncertainty. \citet{Kjeldsen2012} proposed that frequency uncertainty scales with $\sqrt{T^2 + \tau^2}$, where $T$ is the time series duration and $\tau$ is the mode lifetime. The mode lifetime $\tau$ is related to the linewidth $\Gamma$ by $\tau = (\pi \Gamma)^{-1}$.

In principle, the linewidth of oscillation modes should vary smoothly as a function of frequency \citep{Appourchaux2014,Lund2017}. Let us consider fitting an  $\ell=2$  mode with a linewidth $\Gamma_2$ , and an adjacent  $\ell=0$  mode with a linewidth  $\Gamma_0$ . If the result shows  $\Gamma_2 < \Gamma_0$ , this discrepancy may indicate the fitting issue described earlier. To address this, we inflated the uncertainty of the $\ell=2$ mode frequency by a factor of $\sqrt{T^{-2} + (\pi\Gamma_0)^{2}}/ \sqrt{T^{-2} + (\pi\Gamma_2)^{2}}$.

Figure~\ref{fig:RoguePDS} shows the modes overplotted on the PDS, while Fig.~\ref{fig:RogueEchelle} indicates the modes in an \ech diagram. As we consider a substantially evolved red giant, a departure in the power distribution from the approximately vertical ridges is seen. Similarly to Hennes, the frequencies displayed in Table~\ref{tab:FreqAsteroRogue} are corrected when input to BASTA. For Rogue, the correction factor is $1+V_r/c = 0.99998$, based on a $V_r$ of $-6.86$ km/s.

\section{Grid-based modelling of red giants}\label{sec:GridModelling}
The modelling of our two red giants will follow the grid-based modelling approach. Stellar modelling involves many sources of degeneracy between various input parameters (see \citealt{Basu17} for a review). This poses a challenge for so-called forward modelling, where models are iteratively calculated to match the observations. In this scheme, proper exploration of high-dimensional parameter spaces can become challenging, leading to issues stemming from the unresolved degeneracies. The grid-based method avoids this problem by trading it for a computationally expensive approach through the formation of large-scale grids of stellar models. Traditionally, the grid-based modelling approach was applied to examine suites of numerous stars using the same grid, without the need for recalculating the models. Here, we choose to employ the grid-based modelling for in-depth single-star studies (see e.g., \citealt{Stokholm19}; \citealt{Winther23}), warranted by the challenging metal-poor and evolved nature of Hennes and Rogue. The production of representative grids for each of the stars in question is costly in terms of computational resources. Yet, it is critical that they contain stellar models that appropriately span the suitable and often wide parameter space, as it provides insight into the degeneracies in and tentative solutions of the modelling.

The stellar evolution models in this work are computed using GARSTEC \citep{Achim08}. The equation of state employed in the code originates from the OPAL group (\citealt{Rogers96}; \citealt{Rogers02}) and augmented in the low temperature end by that of Mihalas-Hummer-Däppen (\citealt{Dappen88}; \citealt{Hummer88}; \citealt{Mihalas88}; \citealt{Mihalas90}). The treatment of atomic diffusion follows the prescription by \citet{Thoul94}. Various compilations for the opacities are used. For high temperatures it is the ones from OPAL (\citealt{Rogers92}; \citealt{Iglesias96}), while for low temperature the opacities from \citet{Ferguson05} are used. The nuclear reaction rate cross-sections originate from NACRE \citep{Angulo99}, except for the reactions $^{14}\mathrm{N}(\mathrm{p},\gamma)^{15}\mathrm{O}$ and  $^{12}\mathrm{C}(\alpha,\gamma)^{16}\mathrm{O}$ which are from \citet{Formicula04} and \citet{Hammer05}, respectively. As both of our stars have low metallicity, there is a possibility of them being $\alpha$-enhanced, as confirmed by the spectroscopic determinations in Table~\ref{tab:SpecComp}. The stellar abundances used were chosen to be from \citet{Asplund09}, for which versions varying the degree of $\alpha$-enhancement $[\alpha/\mathrm{Fe}]$ in steps of 0.1 dex were calculated and the opacities adjusted. When calculating the synthetic magnitudes of the models, the bolometric corrections by \citet{Hidalgo18} were used.

\subsection{Tailored grids of stellar models}\label{subsec:Grids}
Carrying out detailed grid-based modelling for Hennes and Rogue entails producing stellar grids for each star. The grids are formed by a local grid-building routine\footnote{Named AUbuild; planned for eventual public release} which samples the defined parameter space of initial parameters, e.g. $M_\mathrm{ini}$, according to a quasi-random Sobol sampling (\citealt{Sobol1}; \citealt{Sobol2}; \citealt{Sobol3}; \citealt{Sobol4}; \citealt{Sobol5}; \citealt{Sobol6}). In setting up these grids, only models on the RGB are included as Hennes and Rogue are in this evolutionary state (see Sects. \ref{sec:Intro} and \ref{sec:Asteroseismic}). However, we also performed initial tests using a wider grid containing tracks evolved through the helium flash and onto the core helium burning phase from \citet{Borre22}. In these tests, we fit the same observables as for the modelling results of Hennes and Rogue presented in Sect.~\ref{sec:Results}, however only the radial mode frequencies were taken into account. The test results confirmed RGB membership, with no likely solutions among post-RGB models. Furthermore, these initial tests provide a rough estimate of the suitable parameter space for each star.

\begin{table}[]
\centering
\caption{The parameter space with the varied dimensions of the tailored grids for Hennes and Rogue.}
\begin{tabular}{lcc}
\multicolumn{1}{c}{Stellar parameter} & \multicolumn{1}{c}{Lower bound} & \multicolumn{1}{c}{Upper bound} \\ \hline 
\multicolumn{3}{c}{\textbf{Hennes -- KIC 4671239}} \\ \hline
M $[\mathrm{M}_\odot]$     &  0.70               & 0.90                       \\
\FeH [dex]                    &  -2.9               & -2.3                       \\
\dnu [$\mu$Hz]                &  16                 & 7                          \\
\hline
\multicolumn{3}{c}{\textbf{Rogue -- KIC 7693833}} \\ \hline
M $[\mathrm{M}_\odot]$     &  0.8                & 1.05                       \\
\FeH [dex]                    &  -2.65              & -2.25                      \\
\dnu [$\mu$Hz]                &  5.5                & 2.5                        \\
\hline
\multicolumn{3}{c}{\textbf{Identically varied parameters}} \\ \hline
$f_\mathrm{ov}$               &  0                  & 0.02                       \\
$\alpha_\mathrm{mlt}$         &  1.5                & 2.00                       \\
$Y_\mathrm{ini}$              &  0.246              & 0.280                      \\
$[\alpha/\mathrm{Fe}]$ [dex]  &  0                  & 0.4                        \\
\hline
\end{tabular}
\label{tab:GridParams}
\end{table}
The evolutionary region where models are recorded is defined by an interval in \dnu, the value of which decreases with evolutionary stage. The models are evolved from the pre-main sequence until a certain \dnu (the lower bounds in Table~\ref{tab:GridParams}). The evolution in GARSTEC is then halted and diagnostic checks are performed, before the calculations are restarted with models being recorded until arriving at the value of \dnu indicated by the upper bound in Table~\ref{tab:GridParams}. An overarching prior on the stellar models is a maximum age of 20 Gyr, for which the calculations are stopped for the given track if reached.

The boundaries for the spanned parameter space of the tailored grids are listed in Table~\ref{tab:GridParams}. The parameter space is firstly defined by the stellar mass and metallicity, for which wide boundaries are used that depend on the star in question. For the remaining parameters, the implementation and boundaries used are identical in the two tailored grids. The convective overshooting efficiency $f_\mathrm{ov}$ according to the exponential scheme of \citet{Freytag96} is varied from zero to modestly above the maximal value $f_\mathrm{ov}=0.016$ found through fits to globular clusters \citep{Hjørringaard17}. The $\alpha$-enhancement is sampled from 0 to 0.4 dex. The mixing length parameter $\alpha_\mathrm{mlt}$ according to the mixing-length description by \citet{Kippenhahn13} and initial helium abundance $Y_\mathrm{ini}$ are allowed to vary freely \citep{LiY2024}. A change in the mixing length can cause differences in $T_\mathrm{eff}$ of $\sim 100$ K on the RGB \citep{Cassisi17, Tayar17}. It was thus sampled in a wide range from 1.5 to 2.0. Altering $Y_\mathrm{ini}$ will affect the luminosity of the models. Moreover, models with $Y_\mathrm{ini}$ below the Big Bang nucleosynthesis value may sometimes provide a better fit when considering individual frequencies. As such, we allow for the lower bound to be slightly below this value of $Y_\mathrm{P}=0.24672\pm0.00017$ (\citealt{Planck16}; \citealt{Pitrou18}; \citealt{Cooke18}). This will allow us to check for this tentative situation and add priors subsequently if needed. The effect of mass loss has been omitted in the modelling. Both stars are very low-metallicity, low-mass, first-ascent red giants that are situated below the RGB bump, resulting in a low degree of mass loss \citep{Tailo20, Tailo22}. An integration of the mass loss according to the prescription of \citet{Reimers77}, along the best-fitting track found for the most evolved star, Rogue (see Sect.~\ref{subsec:RogueResults}), finds a cumulated mass loss in the order of $10^{-5}$ M$_\odot$, which we found does not impact the modelling.

The tailored grid for Hennes consists of 4096 stellar tracks, while the grid for Rogue is considerably larger consisting of 8192 stellar tracks. The initial test-fitting of Rogue produced a discrepancy between the observed and modelled temperature of the star. Additionally, the overlap with the magnitudes was at times poor. Lastly, Rogue initially indicated a tentative higher mass solution. The combination of these aspects warranted the larger and more exhaustive grid for Rogue.

\subsubsection{Pulsation code and model frequencies}\label{subsubsec:IndiFreqs}
The individual oscillation frequencies for all models in the grids were computed with the Aarhus Adiabatic Pulsation Package (ADIPLS; \citealt{ADIPLS08}, version 0.4). Importantly, the extension developed by \citet{Larsen24} was implemented. This allows for efficient computation of the individual frequencies in the observable intervals for each model. Previously, the derivation of the individual frequencies for RGB models was only feasible when performing forward-modelling, for which the frequencies could be computed for the few best-fitting models. Here, for the first time in context of grid-based modelling, the mixed modes of both the $\ell=1,2$ individual frequencies are available for all stellar models before the fitting is performed. 

This difference enables us to estimate \dnu for the stellar models according to the individual frequencies instead of Eq.~\ref{eq:DnuScal}. Furthermore, it will allow for the fitting of the observed mixed-mode frequencies of Hennes and Rogue to the grids, which is a crucial extension to prior modelling efforts of evolved red giants.

\begin{figure*}[t]
    \includegraphics[width=17cm]{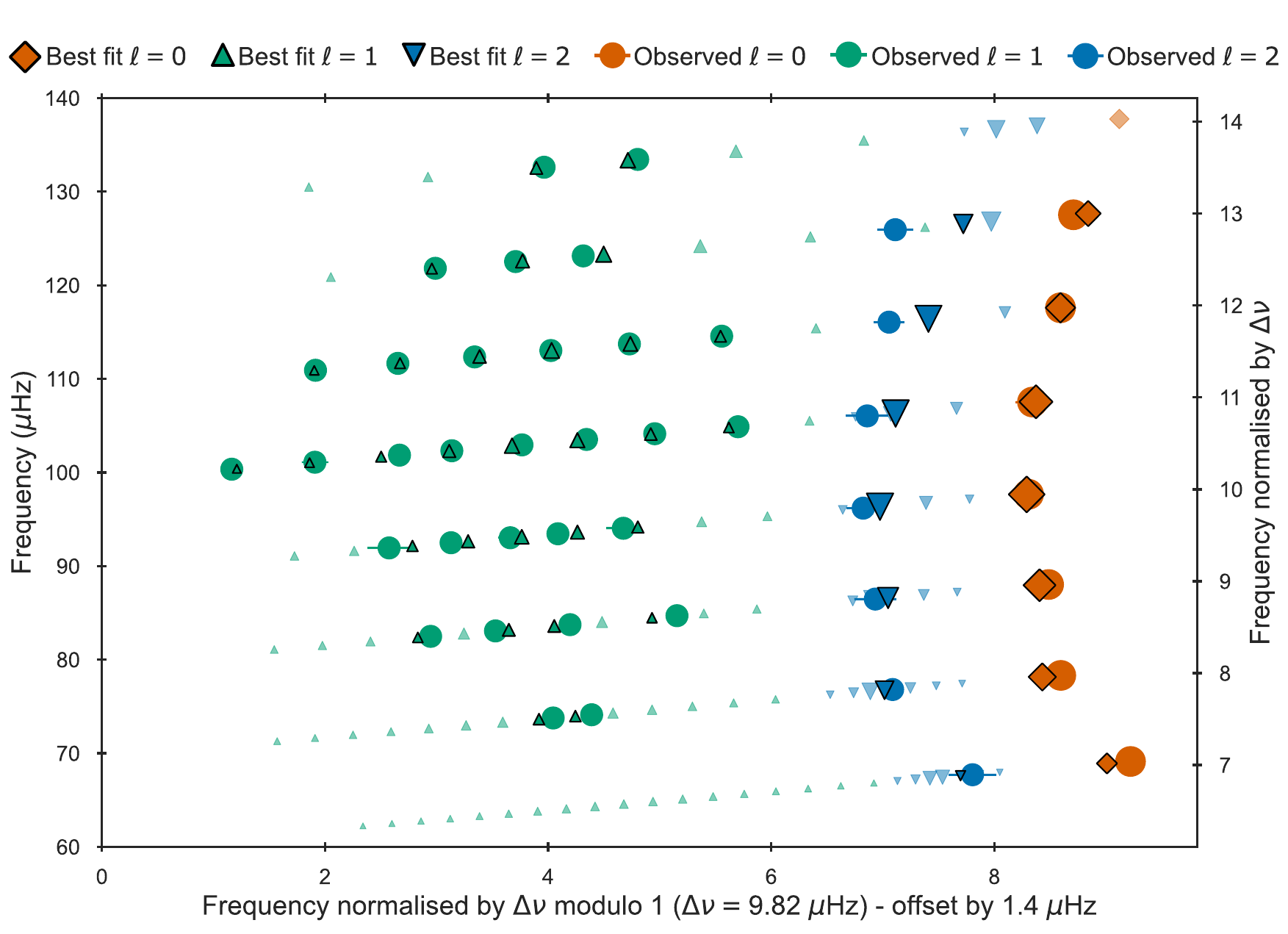}
    \centering
    \caption{Échelle diagram of Hennes with the fitted individual frequencies matched to the modes of the best-fitting model. The orange diamonds, green upwards and blue downwards triangles depict the $\ell=0$, $\ell=1$ and $\ell=2$ modes, respectively. The cubic term of the surface correction by \citet{Ball14} has been applied to the model frequencies, with a surface correction coefficient of $a_3 = -7.0512\cdot 10^{-8}$. The size of the model modes is scaled related to the inverse of their respective inertia. An offset has been applied to the x-axis for improved visualisation to avoid wrapping of the radial ridge}.
    \label{fig:Hennes_freqfit}
\end{figure*}

\subsection{Fitting observations to models}\label{subsec:BASTA}
The fitting algorithm used to match observations to models is the BAyesian STellar Algorithm (BASTA; \citealt{Aguirre22}). It relies on Bayesian inference to infer a set of model properties given the provided observables. It does so through Bayes' theorem, which allows for the combination of prior knowledge on the stellar parameters $\mathbf{\Theta}$ with the data $\mathcal{\mathbf{D}}$ to yield the likelihood $P(\mathcal{\mathbf{D}}|\mathbf{\Theta})$ of observing the data given the model parameters. For specific details see \citet{Aguirre22}. The resulting product from BASTA is a marginalised posterior distribution readily available for inspection. It can be multimodal, meaning it can display more than a single solution for given stellar parameters. As such, is is important to verify the posterior after running the fitting routines. The preferred solution for each parameter is given as the median of the posterior, with uncertainties obtained from the credibility interval formed by the 16th and 84th percentiles of the distribution. Importantly, BASTA is a versatile algorithm that allows for various different observables to be fitted in combination, enabling the detailed modelling that our two stars require.  

For clarity, we summarise how the three global asteroseismic parameters are estimated in our modelling. We define \numax according to the asteroseismic scaling relation in Eq.~\ref{eq:NumaxScal}. Since we have the individual frequencies available, \dnu is determined from weighted differences between the radial frequencies in the given stellar model, following the approach in \citet{White11}. The value of $\Delta\Pi_1$ is determined from asymptotic expressions for a given model (\citealt{Mosser12}, see \citealt{Aguirre22} Sect.~4.1.6 for specific details).

\subsubsection{Fitting individual frequencies}\label{subsubsec:FrequencyMatching}
The fitting of the individual frequencies follows the default algorithm within BASTA. This consists of two steps for each model. Firstly, given the abundance of mixed modes in the models, the matching of observed modes to modes in the model is performed. This is done using the standard method in BASTA (see Sect.~3.1 of \citealt{Ball20}; \citealt{Aguirre22}; Stokholm et al. in prep), with subtle adjustments for application to red-giant stars (a summary of the algorithm can be found in Appendix~\ref{app:E}). Secondly, after identifying the matching $\ell=0,1,2$ model frequencies, they are corrected using the surface correction of \citet{Ball14} using only the cubic term. We choose to use only the cubic term as we fit a modest number of modes for Rogue, which means we are unable to properly constrain the inverse term of \citet{Ball14} and risk overfitting. We note, however, that the posteriors remain unchanged when using the complete formulation. 

There exist previous studies on the application of this surface correction on the RGB, which mention an over-correction issue due to the inertia scaling (\citealt{Li18}; \citealt{Ball18}). The application of the surface correction may result in the non-radial mixed modes being out of order, i.e. the monotonic relationship between mode order and frequency of the theoretical frequencies is broken. This is because the surface correction shifts an acoustic resonance beyond the adjacent oscillations due to its comparably lower inertia. Matching the observed mixed modes to the surface corrected frequencies would thus mean potentially matching to an unphysical representation of the oscillation spectrum. We avoid this as the matching in BASTA is performed before application of the surface correction. Subsequently, when the surface correction is applied, we find that the effect described in \citet{Ball18} impacts the models in the grid for Rogue. Yet, due to the order of the matching procedure, it will not change the statistics of the performed fitting. A discussion regarding the performance of the frequency matching algorithm can be found in Sect.~\ref{subsec:RogueDiscussion}.

\section{Results of asteroseismic modelling}\label{sec:Results}
The modelling of Hennes and Rogue was extensively explored and included a variety of different fitting cases and consistency tests. In the following, for the sake of brevity and clarity, we present only the case that represents the key extension to prior efforts that this work contributes: the fitting of the individual frequencies. As will become clear, this case is the only approach that yields self-consistent results and provide proper overlap between the constraints from the observables. We thus fit the individual frequencies for Hennes and Rogue in Tables \ref{tab:FreqAsteroHennes} and \ref{tab:FreqAsteroRogue}, respectively. In addition the fits included the observed spectroscopic parameters \teff, \FeH and $[\alpha/\mathrm{Fe}]$ along with the distance through the combination of parallax with the \gaia G, BP and RP magnitudes.

\subsection{Hennes -- KIC4671239}\label{subsec:HennesResults}
\begin{figure}[]
    \resizebox{\hsize}{!}{\includegraphics[width=\linewidth]{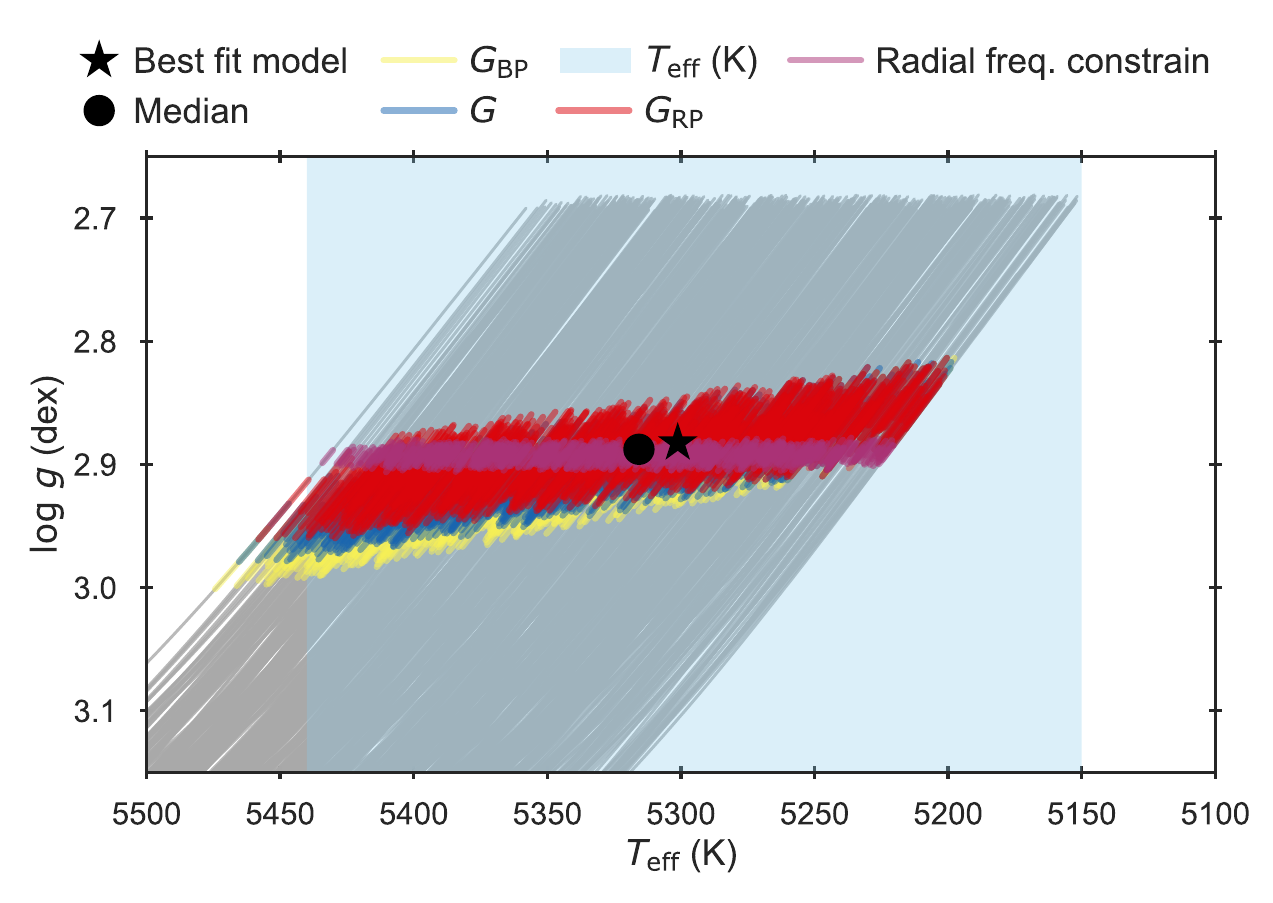}}
    \centering
    \caption{Kiel diagram of Hennes displaying a representative number of stellar tracks in the grid. The observed constraints on the fitted parameters are overlaid to show an overlap with the median and best-fitting model from the posterior.}
    \label{fig:HennesKiel}
    \vspace{-1mm}
\end{figure}

The results for Hennes are displayed in Table~\ref{tab:HennesResults}, compared with the results of \citet{Puls22}. For inspection of the posteriors, see Fig~\ref{fig:HennesCorner}. The detailed asteroseismic modelling with tailored grids and individual frequencies in this work has solved the tension of prior results displaying a low stellar age (presented in Sect.~\ref{sec:Intro}). Hennes being a metal-poor halo star (see Sect.~\ref{subsec:GalaArc}) is now classified as a low-mass high-age star, in coherence with expectations.

\begin{table}
\renewcommand{\arraystretch}{1.3}
\centering
\caption{Modelling results for Hennes. All stellar parameters are given by the median of the posterior distribution from BASTA, with uncertainties as the 16th and 84th quantiles.}
\vspace{-1mm}
\begin{tabular}{lccc}
\multicolumn{3}{c}{\textbf{Hennes -- KIC 4671239}} \\ \hline
\multicolumn{1}{l}{\thead{Stellar parameter}} & \multicolumn{1}{c}{\thead{This work}} & \multicolumn{1}{c}{\thead{Alencastro Puls \\ et al. 2022}} \\ \hline 
$M$ [$M_\odot$]  &  $0.78^{+0.04}_{-0.03}$     &  $1.01^{+0.02}_{-0.02}$   \\
$R$ [$R_\odot$]  &  $5.26^{+0.09}_{-0.07}$     &  $5.70^{+0.05}_{-0.05}$   \\
Age [Gyr]        &  $12.1^{+1.6}_{-1.5}$       &  $5.4^{+0.4}_{-0.3}$      \\
\numax [\muHz]     &  $90.84^{+1.35}_{-1.01}$  &  --    \\
$\dnu$ [\muHz]     &  $9.84^{+0.03}_{-0.03}$   &  --    \\
$\Delta\Pi_1$ [s]   &  $67.37^{+0.56}_{-0.55}$ &  --    \\
\end{tabular}
\label{tab:HennesResults}
\vspace{-5mm}
\end{table}

\begin{figure*}[t]
    \includegraphics[width=17cm]{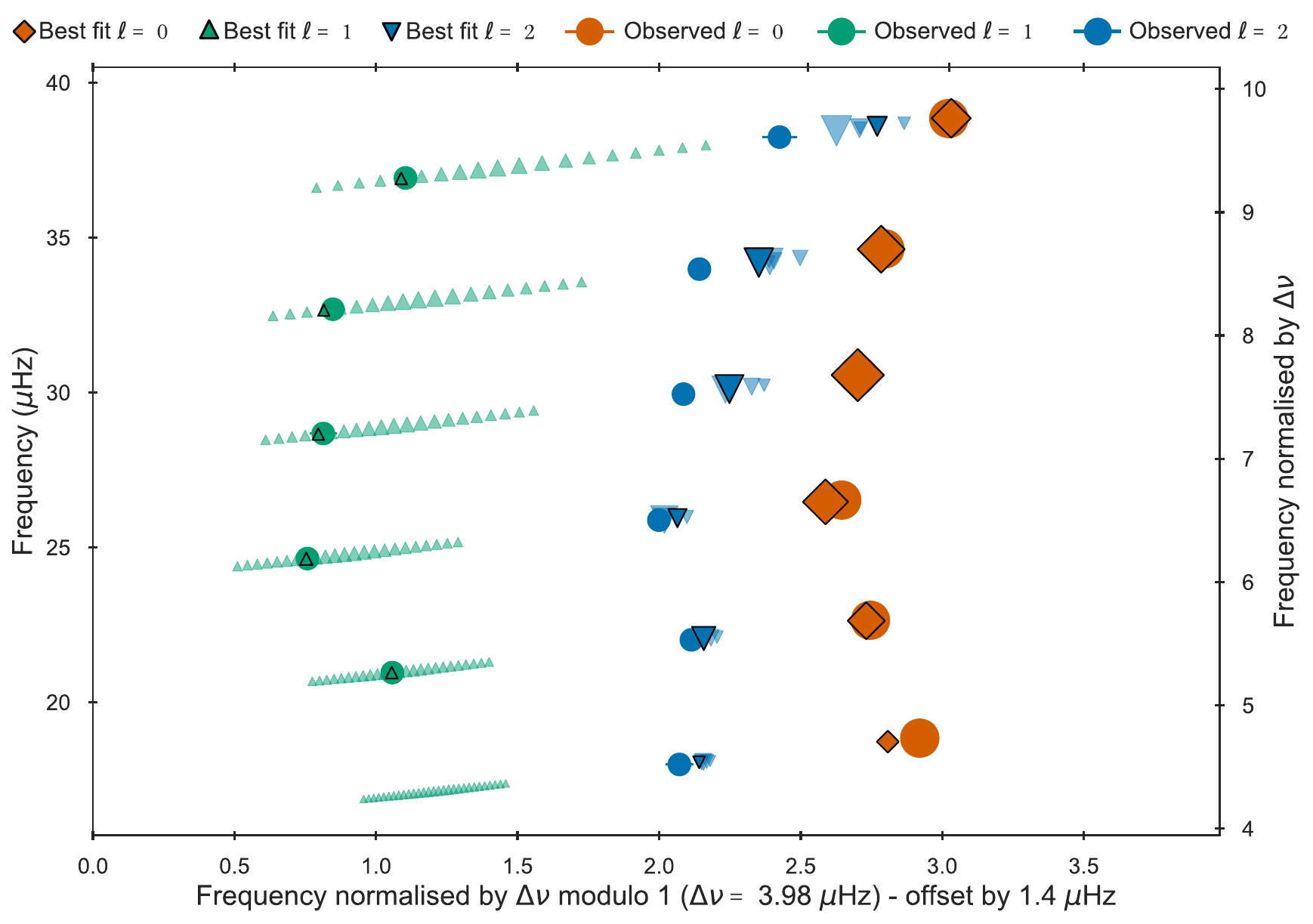}
    \centering
    \caption{Échelle diagram of Rogue with the fitted individual frequencies matched to the modes of the best-fitting model. The orange diamonds, green upwards and blue downwards triangles depict the $\ell=0$, $\ell=1$ and $\ell=2$ modes, respectively. The cubic term of the surface correction by \citet{Ball14} has been applied to the model modes, with a surface correction coefficient of $a_3 = -1.5554\cdot 10^{-7}$. The size of the model modes is scaled related to the inverse of their respective inertia. An offset has been applied to the x-axis for improved visualisation to avoid wrapping of the radial ridge}.
    \label{fig:Rogue_freqfit}
\end{figure*}

Figure~\ref{fig:Hennes_freqfit} show the fit of the individual frequencies in a replicated \ech diagram. The fit to the individual frequencies is satisfactory and shows a best-fitting model with a mixed mode pattern matching that of Hennes very closely. Fitting the dipole mixed-modes directly allows us to capture the underlying period spacing of the model, clearly seen in the most populated mode orders in Fig.~\ref{fig:Hennes_freqfit}. The resulting prediction of $\Delta\Pi_1$ from the fit is also consistent with the observed value. The modelled \dnu value is in accordance with the observed. We note that the three lowest radial modes in the figure deviate from the model due to significant curvature of the observed ridge \citep{Mosser13}. Furthermore, as the cubic surface correction of \citet{Ball14} is scaled with mode inertia, the radial modes are shifted further than the quadrupoles for a given surface correction coefficient. This results in a erroneous appearance of the small spacing $\delta\nu_{02}$ in the surface-corrected model in Fig.~\ref{fig:Hennes_freqfit}, but does not reflect the small spacing in the model. Notably, the predicted \numax value from the fit deviates significantly from the observed. This discrepancy is discussed in detail in Sect.~\ref{subsec:numaxDiscrep}.

Figure~\ref{fig:HennesKiel} illustrates the result in a Kiel diagram (a spectroscopic HR diagram) plotting the surface gravity $\log(g)$ against effective temperature. Only the tracks containing statistically significant models are plotted, forming the dense distribution of stellar tracks seen. The observational constraints for the fitted parameters are overlaid onto the plot for a representative region along each track depicting the 1-$\sigma$ uncertainty \citep{Hjørringaard17}. Notably, the frequency constraint band is formed based on a selection of models matching the radial frequencies, not taking into account the non-radial oscillations in the spectrum. The results display a satisfactory overlap between all observational constraints. Importantly, this kind of overlap remained unattainable in the prior efforts when fitting the global asteroseismic parameters and, crucially, persists in this work if we perform the same fitting approach (see Appendix~\ref{app:C}). Hence, a proper inference for Hennes is only attainable when utilising the individual mode frequencies to avoid the dependence on the asteroseismic scaling relations, as has been presented here.

\subsection{Rogue -- KIC7693833}\label{subsec:RogueResults}
The results obtained for Rogue are displayed in Table~\ref{tab:RogueResults}, where the results of \citet{Pinsonneault24} are also included. For inspection of the posteriors, see Fig.~\ref{fig:RogueCorner}. The obtained parameters in this work indicate a low-mass star with an accordingly high age. These results are in tension with those obtained in prior efforts and with \citet{Pinsonneault24}, yet now comply with the expectations for a low-metallicity star.

\begin{table}
\renewcommand{\arraystretch}{1.3}
\centering
\caption{Modelling results for Rogue. All stellar parameters are given by the median of the posterior distribution from BASTA, with uncertainties as the 16th and 84th quantiles.}
\begin{tabular}{lcc}
\multicolumn{3}{c}{\textbf{Rogue -- KIC 7693833}} \\ \hline
\multicolumn{1}{l}{\thead{Stellar parameter}} & \multicolumn{1}{c}{\thead{This work}} & \multicolumn{1}{c}{\thead{\citet{Pinsonneault24}}} \\ \hline 
$M$ [$M_\odot$]    &  $0.83^{+0.03}_{-0.01}$      & $1.05\pm 0.040$      \\
$R$ [$R_\odot$]    &  $9.53^{+0.14}_{-0.06}$      & $10.32\pm 0.18$   \\
Age [Gyr]          &  $10.3^{+0.6}_{-1.4}$        & $4.85^{+0.60}_{-0.52}$   \\
\numax [\muHz]     &  $29.83^{+0.27}_{-0.23}$    & --       \\
$\dnu$ [\muHz]     &  $4.072^{+0.01}_{-0.004}$   & --       \\
$\Delta\Pi_1$ [s]    &  $53.62^{+0.32}_{-0.99}$    & --       \\
\end{tabular}
\label{tab:RogueResults}
\end{table}

\begin{figure}[]
    \resizebox{\hsize}{!}{\includegraphics[width=\linewidth]{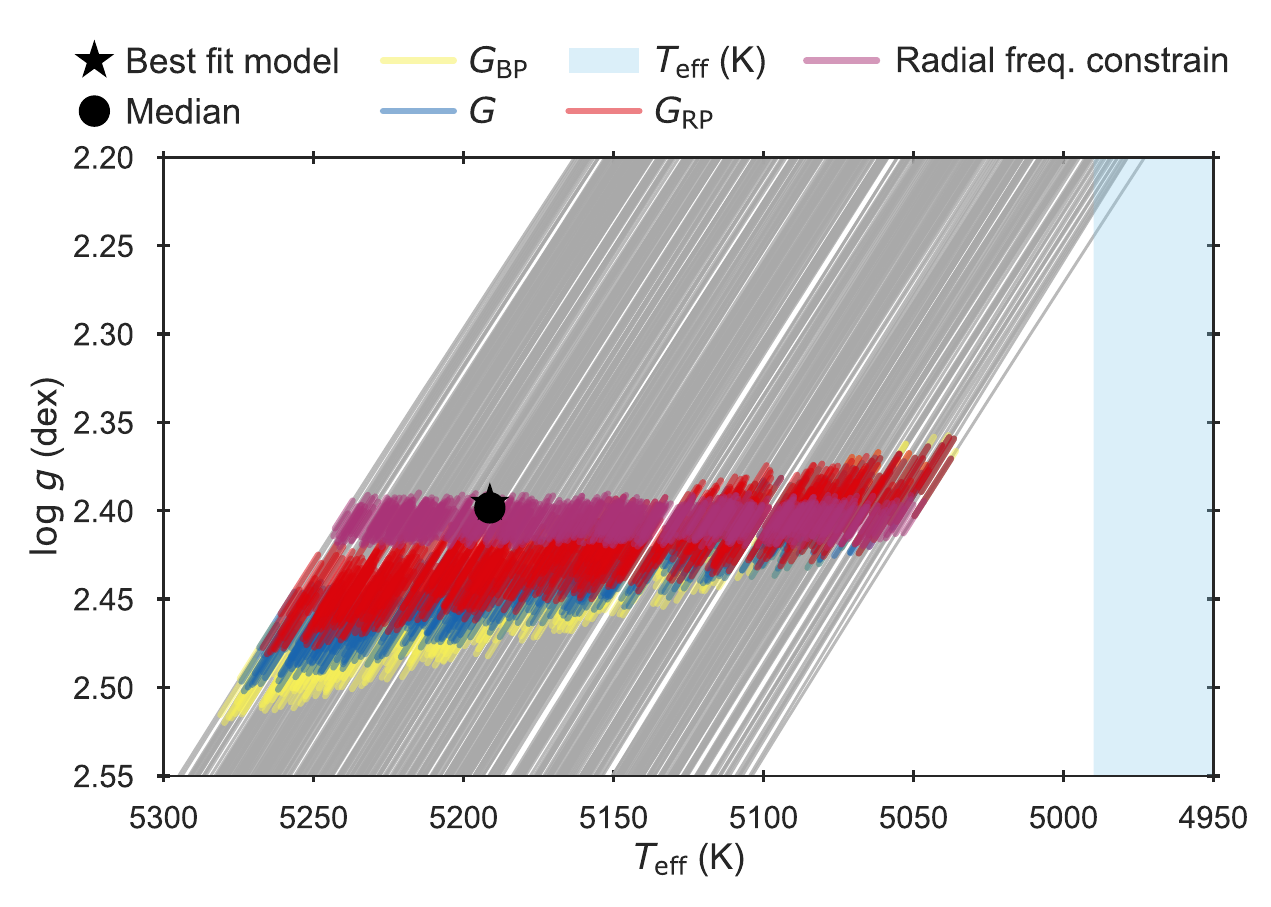}}
    \centering
    \caption{Kiel diagram of Rogue displaying a representative number of stellar tracks in the grid. The observed constraints on the fitted parameters are overlaid to show an overlap with the median and best-fitting model from the posterior. A clear discrepancy between the observed and modelled temperature is found, with the lower 1-$\sigma$ uncertainty visualised by the boundary of the blue region.}
    \label{fig:RogueKiel}
\end{figure}

Figure~\ref{fig:Rogue_freqfit} displays the fit to the individual frequencies for Rogue. The more evolved nature of the best-fitting model is immediately apparent from the denser theoretical frequency spacing for each acoustic mode order. The fit to the individual frequencies is satisfying, however with some artefacts of the frequency matching algorithm impacting the quadrupole modes. This will be further discussed in Sect.~\ref{subsec:RogueDiscussion}. The modelled \dnu value is slightly higher than the observed. This is due to the \dnu value in the grid being estimated from a weighted average between the non-surface-corrected radial frequencies. Hence, the obtained value will be somewhat higher, but by inspection of Fig.~\ref{fig:Rogue_freqfit} is suitable for creating the vertical ridge structure in the \ech diagram. It may be noted that the modelled period spacing of $\Delta\Pi_1 = 53.6^{+0.3}_{-1.0}$ s is roughly in accordance with the uncertain observational estimate of $\Delta\Pi_1 = 56.6$ s discussed in Sect.~\ref{subsec:AsteroRogue}. The \numax discrepancy found for Hennes is also clear for Rogue. 

Figure~\ref{fig:RogueKiel} illustrates the result in a Kiel diagram, similar to that made for Hennes. At first glance, it appears counter-intuitive why the best-fitting model and median posterior solutions do not lie at lower temperatures where the frequency constraints overlap perfectly with the Gaia magnitudes. This is because the frequency constraint band is formed based on the radial frequencies only, allowing the best-fitting model and median solutions, which take into account the complete mode spectrum and other fit parameters, to lie away from the visual overlap. A clear $\sim$ 2.3 $\sigma$ deviation from the observed temperature is found, an unresolved complication present for Rogue. Crucially, the best-fitting model and seismic solution is consistent with all other observables, and provide a modelling result, which is not in tension with prior expectations from the low-metallicity nature of the star. In Sect.~\ref{subsec:RogueDiscussion} we discuss the nuances of the modelling challenges for Rogue, where a more thorough evaluation will be presented. 
\section{Asteroseismic characterisation from individual frequency modelling}\label{sec:Discussion}
This work has demonstrated that rigorous asteroseismic modelling is feasible for evolved red giants, even in the most challenging cases of the very low-metallicity regime. We emphasize that the results obtained are not reproduced if one considers the global asteroseismic parameters in the fitting (see Appendix \ref{app:C}). This suggests that future studies of red giants could benefit from relying more on the individual frequencies as direct observable constraints in the modelling.

Hennes is the first metal-poor star with \numax below 100 \muHz to be asteroseismically modelled with $\ell=0,1,2$ mixed-mode individual frequencies within a grid-based methodology. Importantly, at this stage of evolution, we can clearly observe several dipole mixed modes within each acoustic mode order. As shown in Fig.~\ref{fig:Hennes_freqfit}, this allows us to capture the underlying period spacing during the modelling, effectively providing constraints on the interior stellar structure. Rogue further extends the coverage of metal-poor stars with asteroseismic individual frequency modelling to the region where \numax is below 35 \muHz. Here, the mixed mode pattern has become so dense and the period spacing so low that we cannot confidently resolve the dipole mixed modes observationally. Additionally, the inertia of the g-dominated modes increases, making their detection difficult. Only the observational estimate for the acoustic resonances of each mode order were fitted; however to models containing all potentially observable mixed modes, which allows for the opportunity of matching several theoretical frequencies. This allowed for the classification of a particularly vexing star; however, a significant temperature discrepancy between the observed and modelled values persists.

\subsection{Modelling nuances of Rogue}\label{subsec:RogueDiscussion}
In Fig.~\ref{fig:RogueKiel} a clear discrepancy between the observed and modelled temperature was seen. A likely cause was mentioned in Sect.~\ref{subsec:Atmospheric}, namely the application of 1D NLTE spectroscopic reduction methods for the very low-metallicity regime. The degeneracy between the temperature and metallicity, as well as the implementation of the microturbulence as a fitting parameter, may very well lead to inconsistencies on the observational side. Furthermore, the temperature of the RGB stellar models is also a topic of discussion. \citet{Cassisi17} showed how the temperature of RGB models is sensitive to $\alpha_\mathrm{MLT}$ and the outer boundary condition employed in the $T(\tau)$ relation for the choice of atmosphere. These aspects were further outlined by \citet{Tayar17}, who presented a metallicity dependent temperature offset for RGB stars when compared to theoretical predictions, suggesting to make the convective mixing length metallicity dependent to account for the effect. \citet{Salaris18} later reanalysed the sample from \citet{Tayar17}, and found that the discrepancy was only significant for $\alpha$-enhanced RGB stars. \citet{Schonhut-Stasik24} also argued that the choice of temperature scales for low-metallicity red-giant models may lead to discrepancies. It is noteworthy that this temperature discrepancy was not present for Hennes while being so prominent for Rogue. A planned future study employing a larger sample of low-metallicity giants will likely aid in classifying the occurrence and behavioural trends of the temperature discrepancy.

In the frequency fit of Rogue in Fig.~\ref{fig:Rogue_freqfit}, certain quadrupole modes appear to be mismatched, for example in the highest mode order. The frequency matching algorithm in BASTA was developed for main-sequence and sub-giant stars (see e.g. \citealt{Stokholm19}). For the former, only one mode exists per acoustic mode order and for the latter there is a large difference in mode inertia between the few modes in each mode order. On the RGB and with the oscillation spectra obtained through the application of the truncated scanning method of \citet{Larsen24}, we obtain a modest number of mixed modes closely spaced in frequency and of comparable inertia within each mode order. This means that the matching algorithm described briefly in Sect.~\ref{subsubsec:FrequencyMatching} may not be entirely suitable for application to red giants. However, we stress that forcing the solution to the "by-eye" optimal solution for the best-fitting model in Fig.~\ref{fig:Rogue_freqfit} has no impact on the choice of optimal model during the fitting. A test was carried out simply matching the observed mode of each degree to the respective model frequency with the lowest inertia within each acoustic mode order. This does not significantly change the drawn posteriors, as the contribution to the likelihood from the above change is less significant than the evolution of particularly the radial frequency spectrum between consecutive models. A proper evaluation of a refined RGB mixed-mode frequency matching algorithm requires statistics on the performance based on more than the two stars at hand, which is the subject of a future study.

\subsection{Circumventing the \numax scaling relation dependence}\label{subsec:numaxDiscrep}
The estimates for \numax obtained from the modelling posteriors showed a significant discrepancy to the observed for both stars. To reiterate, the \numax value of the models are predicted using the scaling relation in Eq.~\ref{eq:NumaxScal} as no other alternative exists. We circumvented the dependence on the scaling relations entirely by fitting the individual frequencies, resulting in the highest-likelihood models being selected based on the direct constraints that the frequencies provide to the interior model structures. This difference suggests a metallicity dependence on the \numax scaling relation, which becomes significant for very low-metallicity stars. This agrees with the dependence of \numax on the Mach number found by \citet{Belkacem11} and with other previous studies finding the same indication, albeit in different metallicity or evolutionary regimes (\citealt{Epstein14}; \citealt{LiT22}; \citealt{Campante23}; \citealt{LiY2024}).

Investigating this metallicity dependence means evaluating a correction factor to the scaling relation, 
\begin{align}
    \numax &= f_{\nu_\mathrm{max}} \nu_\mathrm{max, scal} \simeq f_{\nu_\mathrm{max}} \left( \frac{M}{M_{\odot}}\right)\left(\frac{R}{R_{\odot}}\right)^{-2} \left( \frac{\teff}{T_{\mathrm{eff},\odot}}\right)^{-1/2}\nu_{\mathrm{max},\odot} \ ,    \\
    f_{\nu_\mathrm{max}} &= \frac{\nu_\mathrm{max, obs}}{\nu_\mathrm{max, BFM}}.
\end{align}
Here, $\nu_\mathrm{max, BFM}$ can be extracted from the best-fitting model of Hennes and Rogue while $\nu_\mathrm{max, obs}$ are the observed values shown in Table~\ref{tab:GlobalAsteroCombined}. The values are $\nu_\mathrm{max, BFM}=89.83$ and $\nu_\mathrm{max, BFM}=29.60$ \muHz, resulting in correction factors of $f_{\nu_\mathrm{max}}=1.101$ and $f_{\nu_\mathrm{max}}= 1.098$, for Hennes and Rogue, respectively. This yields a difference in \numax of $\sim10\%$, showing that the uncorrected \numax scaling relation in the metal-poor regime is significantly less accurate than near solar metallicities. We can express this difference in mass using the observed \numax, \dnu and \teff to calculate the scaling relation estimate as,
\begin{align}
        M &= \left(\frac{\nu_\mathrm{max, obs}}{\nu_\mathrm{max,\odot}}\right)^3 \left(\frac{\dnu}{\Delta\nu_\odot}\right)^{-4} \left(\frac{\teff}{T_\mathrm{eff,\odot}} \right)^{-1/2}, \label{eq:ScalMass}\\
         &= f_{\nu_\mathrm{max}}^3 \left(\frac{\nu_\mathrm{max, BFM}}{\nu_\mathrm{max,\odot}}\right)^3 \left(\frac{\dnu}{\Delta\nu_\odot}\right)^{-4} \left(\frac{\teff}{T_\mathrm{eff,\odot}} \right)^{-1/2}.
\end{align}
Using the scaling relation in Eq.~\ref{eq:ScalMass}, this results in a $\sim30\%$ and $\sim 40\%$ larger mass from the scaling relation than from the determination when performing asteroseismic individual frequency modelling. This seconds the findings of \citet{Huber24} who similarly recovered a $\sim 24\%$ difference in mass for KIC 8144907 with $\FeH=-2.66$ dex. The tension remains when using proposed \dnu metallicity corrections \citep{Viani17} and repeating the above exercise. For these three RGB stars with individual frequency modelling, the situation thus hints towards an increasing discrepancy with evolution for these similarly very metal-poor giants. 

Lastly, we note that for the stars KIC 7341231 \citep{Deheuvels12}, $\nu$ Indi \citep{Chaplin20} and HD 140283 (Lundkvist et al. in prep) with individual frequency modelling shown overplotted in Fig.~\ref{fig:Overview}, the tentative metallicity dependence on \numax between observed and model values was also seen.

\subsection{Enhanced mode mixing at low metallicity}\label{subsec:InteriorMixing}
During this work concerning stellar modelling of low-metallicity giants, a property of the interior profiles in the models was noticed. The extent of the evanescent region was increased and the coupling weakened with increasing metallicity. Here, we wish to briefly clarify and present this trend seen in the models. 

\begin{figure}[]
    \resizebox{\hsize}{!}{\includegraphics[width=\linewidth]{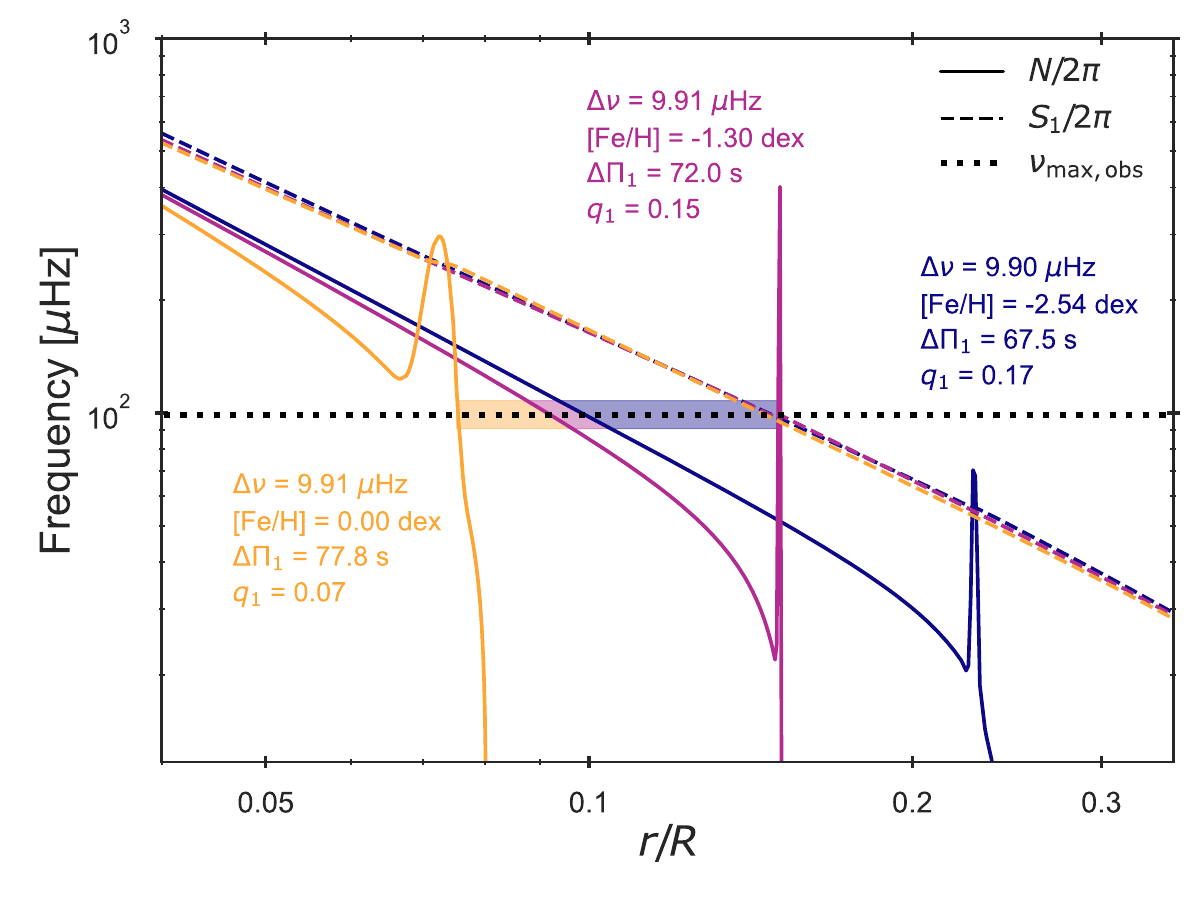}}
    \centering
    \caption{Propagation diagram for three stellar models depicting the characteristic frequencies in the interior near the lower convective boundary. The best-fitting model of Hennes is plotted alongside two models of similar evolutionary state, i.e. similar \dnu value, with identical initial parameters except for a variation in \FeH. The shaded region around the observed \numax of Hennes represents the extent of the evanescent region for the three models, which increases with metallicity.}
    \label{fig:HennesBFMvsFeH}
\end{figure}
\begin{figure*}
    \centering
    \includegraphics[width=17cm]{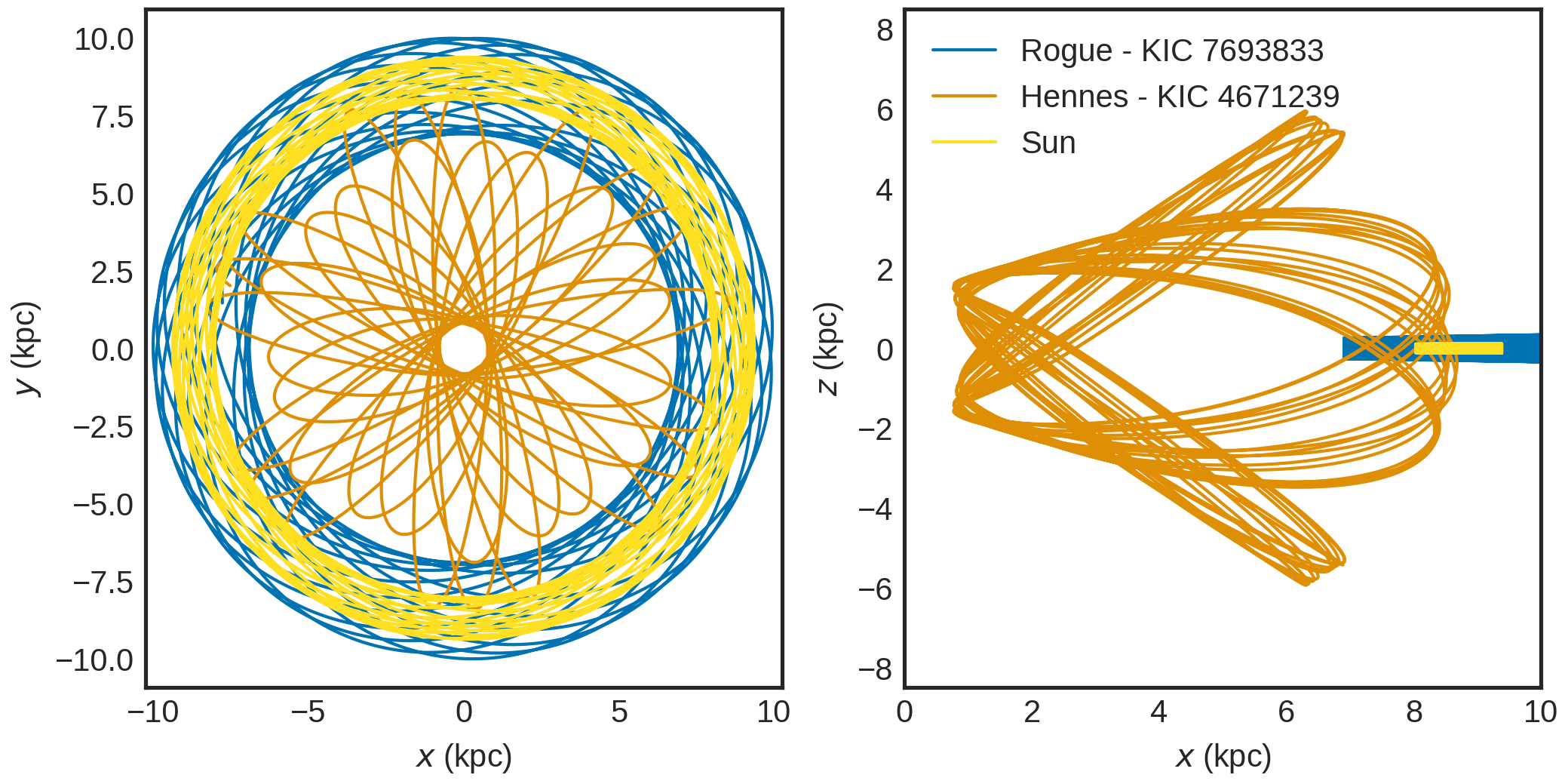}
    \caption{The integrated orbit of \rogue and \hennes within the Milky Way's potential over a $10$~Gyr period, depicted in Cartesian Galactocentric coordinates. The left panel presents a top-down view of the Galaxy ($x, y$), while the right panel shows an edge-on perspective ($x, z$). The Sun's integrated orbit is included in both views as a reference.}
    \label{fig:orbitintegration}
\end{figure*}

Figure~\ref{fig:HennesBFMvsFeH} depicts a propagation diagram of three stellar models, showing the proximity of the oscillation cavities in the deep interior of the models. The very low-metallicity model is the best-fitting model of Hennes. The other two were calculated with identical initial parameters and varying the metallicity to $\FeH=-1.3$ and $\FeH = 0.0$. The buoyancy frequency $N$ \citep{Tassoul80} and Lamb frequency $S_\ell$ \citep{Lamb32} are calculated using Eqs.~2 and 3 of \citet{Larsen24}, respectively. The figure shows that the depth of the convective zone increases with increasing metallicity, as a result of the dependence of convection on the opacity and thus chemical composition. This effectively increases the extent of the evanescent region (as shown by the shaded bands of each model) for a representative frequency here chosen as the observed \numax of Hennes. Equations 11 and 12 of \citet{Hekker17} allows us to evaluate the dipole period spacing and Eq. 9 of \citet{Larsen24} provides an estimate of the coupling constant. The obtained values, overplotted on Fig.~\ref{fig:HennesBFMvsFeH}, clearly illustrate the same trend: a decreasing period spacing and increasing coupling constant with decreasing metallicity.

This effect should lead to a stronger mixing between the p- and g-modes of low-metallicity stars, resulting in a larger frequency displacement for the mixed modes within a given acoustic order. Additionally, the difference in the period spacing and coupling constant will affect Eq. 28 of \citet{Mosser18}, which describes the frequency range in which mixed modes should be observable around an acoustic resonance, hinting towards a possible metallicity dependence. Confirming the postulated property of mixed-mode behaviour as a function of metallicity would require a dedicated study of the trend in observed oscillation spectra of RGB stars, which is beyond the scope of this work.

\subsection{Galactic context}\label{subsec:GalaArc}
The Galactic orbital characteristics of the stars were determined from the 5D astrometric information and line-of-sight velocities from \gaia DR3 \citep{Gaia16, GaiaDR3}.
We use the \textsc{Python} package \texttt{galpy} \citep{galpy} for this computation of the Galactic orbital properties. As a description of the Milky Way potential, we use the axisymmetric gravitational potential \texttt{McMillan2017}  \citep{mcmillan2017}. As for the Galactic location and velocity of the Sun, we assume  $(X_{\odot},Y_{\odot},Z_{\odot})=(8.2,0,0.0208)$~\si{\kilo pc} and  $(U_{\odot},V_{\odot},W_{\odot})=(11.1,12.24,7.25)$~\si{\kilo\metre\per\second} with a circular velocity of \SI{240}{\kilo\metre\per\second} \citep{schonrich2010,bennett2019,gravity2019}.

We use the implementation of the action-angle estimation algorithm \emph{St\"ackel fudge} \citep{binney2012} in \texttt{galpy} with a focal length focus of $0.45$ to calculate orbit information such as actions, eccentricity, and maximum orbit Galactocentric height.
By leveraging the uncertainties and correlations in the astrometric measurements to create a multivariate normal distribution, we perform 10,000 iterations to generate distributions of the orbital properties. The 16th, 50th, and 84th percentiles of these distributions are then used as numerical estimates of the orbital values and their associated uncertainties. The complete list of orbital properties for \hennes and \rogue can be found in Table~\ref{tab:orbitalparams} in the Appendix.

Fig.~\ref{fig:orbitintegration} illustrates the integrated orbits of \hennes and \rogue within the Milky Way's gravitational potential, highlighting their differing extents. \hennes traverses the dense inner regions of the Galaxy, with an inner turning point nearly 1 kpc from the Galactic centre. Additionally, \hennes frequently moves far from the Galactic disk, reaching heights several kiloparsecs above the Galactic midplane. In contrast, while \rogue does venture beyond the Local Solar Neighbourhood, its orbit more closely resembles that of the Sun, differing significantly from the more extensive and dynamic path of \hennes.

\begin{figure}
    \centering
    \resizebox{\hsize}{!}{\includegraphics[width=\linewidth]{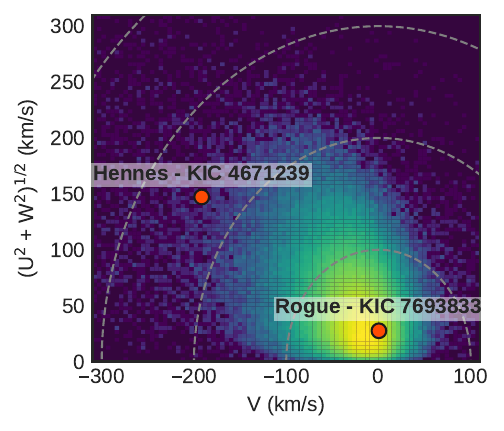}}
    \caption{A Toomre diagram showing the Galactic velocity components,  with the vertical axis representing motion perpendicular to the plane and the horizontal axis depicting motion within the plane. The location of \hennes and \rogue in this space are marked with red dots. The background density map represents the number count of single targets from \gaia~DR3 with reliable astrometric data and available line-of-sight velocities (\texttt{astrometric\_params\_solved = 95}, \texttt{non\_single\_star = 0}, \texttt{ruwe < 1.4}, and \texttt{rv\_nb\_transits > 0}).}
    \label{fig:toomre}
\end{figure}

Fig.~\ref{fig:toomre} shows the location of the stars in a Toomre diagram, in which the horizontal axis depicts motion around in the plane and the vertical axis represents motion perpendicular to the horizontal axis. This representation helps distinguish different Galactic components more clearly. In this context, \hennes stands out as behaving like a typical halo star, characterized by its lower rotational velocity within the Galactic plane and higher velocities in the vertical (up-down) and radial (in-out) directions compared to the Sun.
\rogue is more intriguing, as it appears to align with the characteristics of a thin disk star, exhibiting only slightly higher velocities in the direction perpendicular to the Galactic plane.

\subsubsection{Possible Galactic origins}
In this section, we assess the possible Galactic origins of the two stars based on their combined kinematics, spectroscopic and age signatures.

Kinematically \hennes looks like a characteristic halo star. It has a low orbital energy and the orbit looks kinematically heated in velocities, eccentricity, and angular momentum. The position of \hennes in angular momenta-orbital energy dimensions could place it in either L-RL3 \citep{ruizlara2022,dodd2023}, Heracles \citep{horta2021}, or in a metal-poor tail of Gaia-Enceladus \citep{helmi2018}. In either of these cases, \hennes would be the most metal-poor member of each association discovered to date.

The orbit of \rogue is similar to those of the older members of the Galactic thin disk, which is peculiar given its age and metallicity. This suggests that \rogue is part of this controversial, ancient, and very metal-poor stellar population in the Milky Way. The discovery of very metal-poor stars exhibiting disc-like kinematics and a preference for prograde over retrograde orbits has recently generated significant interest within the community \citep{sestito2019, cordoni2021,belokurov2022,carollo2023,bellazzini2024,ardernarentsen2024,fernandezalvar2024}. Studies of these metal-poor stars with prograde planar orbits can provide us insights into the earliest phase of the formation of the Milky Way discs and there is still many unanswered questions. From simulations of model galaxies, \citet{sotilloramos2023} found that the fraction of very metal-poor stars in the kinematically cold disc component varies from $5-10$~\% up to $40$~\%, with typical values around $20$~\% and an occurrence rate that seem to decrease with decreasing metallicity. Other comparisons with simulations of Milky Way analogues suggests that most stars in this population likely originated from accretion events on nearly planar orbits \citep{sestito2021} as the emergence of this population is unlikely to be induced by a rotating galactic bar \citep{yuan2024}. \citet{nepal2024} emphasized that many disc galaxies at high redshifts possess an ancient and kinematically cold disc formed in situ, suggesting that this metal-poor population could represent our Galaxy's analogue.

If \rogue is representative of the population, then we can use its stellar age as a constraint when assessing the possible origins. Given \rogue's characteristics, the detailed modelling of Rogue using asteroseismic constraints is more aligned with the view that this population originates from multiple accretion events, in which the accretion occurred close to planar orbits, than with the view that this stellar population can solely be a single, unperturbed ancient metal-poor thin disk. As it is possible that multiple channels contribute to the formation of this population, with \rogue belonging to one and not the other, a future study employing asteroseismology for this population of stars appears to be the next step.

\section{Conclusion}\label{sec:Conclusion}
This work proves that asteroseismic individual frequency modelling is possible within the grid-based modelling scheme for evolved red giants. Proper determination of global stellar properties in the very metal-poor regime has been shown to crucially depend on how the stellar inference is made. The main conclusions of this paper are as follows: 
\begin{itemize}
    \item The two very low-metallicity stars, KIC 4671239 and KIC 7693833, have been characterised with stellar properties that are coherent with expectations from Galactic archaeology. 
    Hennes and Rogue were found to have masses of $0.78^{+0.04}_{-0.03}$ and $0.83^{+0.03}_{-0.01} \ M_{\odot}$ with an age of $12.1^{+1.6}_{-1.5}$ and $10.3^{+0.6}_{-1.4}$ Gyr, respectively. Both stars had for a long time defied such characterisation, as they are situated in a very difficult region of parameter space for stellar modelling, yet individual frequency modelling in combination with detailed grid-based approaches has made it possible. Crucially, the obtained self-consistency could not be obtained when fitting the global asteroseismic parameters in the past nor in the contemporary work. This fact speaks to the benefits of future application of individual frequency modelling for giants. 
    \item A metallicity dependence of the asteroseismic scaling relation for \numax was seen through a clear discrepancy between the observed and modelled values for Hennes and Rogue of ${\sim}10\%$, with the \numax predicted from the asteroseismic scaling relation being too small compared to the observations. In the determination of global stellar properties for very metal-poor stars in general, such a difference would correspond to overestimations in mass by ${\sim}30\%$, leading to wrongful conclusions about the age demographics of the population. This result is consistent with prior findings by e.g. \citet{Belkacem11}, \citet{LiT22} and \citet{Huber24}. Yet, it is also in slight tension with studies by \citet{Zhou24}, finding that for the case of main sequence stars \numax shows no metallicity dependence. However, the conclusion of \citet{Zhou24} may not hold in the case of more evolved stars in the sub-giant or red-giant branch phases of their lives. To answer this question definitively will require further studies on both the theoretical and observational side, something which is currently in progress.
    \item Accurate and precise stellar ages are crucial for unravelling the timeline of events in the Galaxy. The impact of the systematic effect of metallicity on the \numax scaling relation can be significant for Galactic archaeology as asteroseismic targets are typically used as calibrators. In recent years, stellar remnants of galaxies that merged with the early Milky Way have been a particular topic of interest for the community \citep[see e.g.][for a review]{helmi2020}. We have identified Hennes to be a member of one of such halo populations, making it the most metal-poor member with detailed modelling using asteroseismic constraints. Rogue seems to be a member of the ancient very metal-poor population with prograde, planar orbits; providing a valuable age constraint to studies of the earliest phase of the formation of the Milky Way disc structures. 
\end{itemize}

We emphasise the possibilities of circumventing the dependence on the asteroseismic scaling relations and directly fitting the individual frequencies of red giants. Observationally, \emph{Kepler} and the work by observational astronomers has supplied us with high-quality detection of the individual frequencies of giants, which is continuously being expanded by missions such as the Transiting Exoplanet Survey Satellite (TESS; \citealt{Ricker14}) and in the future the PLAnetary Transits and Oscillations of stars mission (PLATO; \citealt{PLATO24}). It is now possible following the work by \citet{Larsen24} to utilise the individual frequencies to their full potential and this paper has demonstrated that it works in one of the most challenging regions for stellar modelling. We therefore suggest that such efforts should be pursued in the future as it may allow for refinement of our understanding of stars on the RGB. For example, investigations resolving the convective zone boundaries probed by the period spacing $\Delta\Pi_1$ could be further developed by the inclusion of the mixed-mode dipole individual frequencies, which directly probes this region in the stellar interior through their sensitivity to the mode coupling (\citealt{Jiang14}; \citealt{Pincon19}). These efforts in turn link to the study of the origin behind magnetic fields in red giant stars (see e.g., \citealt{Deheuvels23}; \citealt{Das24}); a field of research still in its infancy.

\begin{acknowledgements}
    We thank the anonymous referee for the detailed and constructive feedback on the manuscript. JRL wishes to thank the members of SAC and the collaborators of the prior efforts for valuable comments and discussions regarding the paper. Furthermore, we appreciate being given access to the data for and the allowed inclusion of Fig.~\ref{fig:Overview} by Daniel Huber. 
    \\ 
    This work was supported by a research grant (42101) from VILLUM FONDEN. Funding for the Stellar Astrophysics Centre was provided by The Danish National Research Foundation (grant agreement no.: DNRF106). MSL acknowledges support from The Independent Research Fund Denmark's Inge Lehmann  program (grant  agreement  no.:  1131-00014B). The numerical results presented in this work were obtained at the Centre for Scientific Computing, Aarhus \url{https://phys.au.dk/forskning/faciliteter/cscaa/}. AS acknowledges support from the European Research Council Consolidator Grant funding scheme (project ASTEROCHRONOMETRY, G.A. n. 772293, http://www.asterochronometry.eu).\\
    This work presents results from the European Space Agency (ESA) space mission Gaia. Gaia data are being processed by the Gaia Data Processing and Analysis Consortium (DPAC). Funding for the DPAC is provided by national institutions, in particular the institutions participating in the Gaia MultiLateral Agreement (MLA). The Gaia mission website is \url{https://www.cosmos.esa.int/gaia}. The Gaia archive website is \url{https://archives.esac.esa.int/gaia}.
    \\
    This paper includes data collected by the Kepler mission and obtained from the MAST data archive at the Space Telescope Science Institute (STScI). Funding for the Kepler mission is provided by the NASA Science Mission Directorate. STScI is operated by the Association of Universities for Research in Astronomy, Inc., under NASA contract NAS 5–26555.   
    \\
    SH acknowledges support from the European Research Council under the European Community’s Horizon 2020 Framework/ERC grant agreement no 101000296 (DipolarSounds).
    \\
    D.S. is supported by the Australian Research Council (DP190100666).
\end{acknowledgements}

\section*{Data availability}
All data and stellar grid products are available upon reasonable request to the first author.

\bibliographystyle{aa.bst} 
\bibliography{bibliography.bib} 

\newpage

\begin{appendix}
\section{Independent peakbagging}\label{app:A}
\begin{table*}[htbp]
\centering
\caption{Estimates of the observed global asteroseismic parameters for \hennes and \rogue from the independent determinations. If no uncertainty was returned, it has been left unspecified. Additionally, if an estimate of the coupling constant $q$ was returned, it is also given.}
\begin{tabular}{lcccc}
\multicolumn{5}{c}{\textbf{Hennes -- \kichennes}} \\ \hline
\multicolumn{1}{c}{Collaborator} & \numax [\muHz] & \dnu [\muHz] & $\Delta\Pi_1$ [s] & $q$ \\ \hline
Sect.~A.2  & $99.87$       & $9.77$        & $62.3$   & -- \\
Sect.~A.3  & --            & $9.798 \pm 0.001$        & $66.53\pm0.01$  & $0.27\pm 0.02$ \\
Sect.~A.5  & $105.6\pm1.1$ & $9.83\pm0.05$ & $66.4\pm1.2$ & $0.26\pm0.03$ \\
\\
\multicolumn{5}{c}{\textbf{Rogue -- \kicrogue}} \\ \hline
\multicolumn{1}{c}{Independent Determination} & \numax (\muHz) & \dnu (\muHz) & $\Delta\Pi_1$ (s) & $q$ \\ \hline
Sect.~A.2  & $32.35$       & $4.06$         & --         & -- \\
Sect.~A.3  & --            & $4.10$         & $105.2\pm 28.0$         & $0.09\pm0.07 $ \\
Sect.~A.5 & $33.2\pm2$    & $4.04\pm0.035$ & $61.3\pm58.1$ & $0.03\pm0.02$ \\
\hline
\end{tabular}
\label{tab:ColabGlobParams}
\end{table*}

This appendix briefly summarises the independent peakbagging performed by the collaborators, and displays the consolidated lists of individual frequencies used in the modelling. Furthermore, Table~\ref{tab:ColabGlobParams} display the available estimates for various global asteroseismic parameters obtained by the collaborators. Note that for \hennes, an alternate measure for the period spacing of $\Delta\Pi_1 = 67^{+0.03}_{-0.16}$~s was determined based on the re-formulation of mixed modes by \citet{Jiang14} and implemented by \citet{Hekker18}, which does not take buoyancy glitches into account.

\subsection{Peakbagging following Li et al. 2020}\label{subapp:Yaguang}
We first identified oscillation modes from regular \'echelle diagrams for p modes as well as stretched \'echelle diagrams for mixed modes. Then using these initial guesses, we extracted the mode frequencies by fitting the power spectrum using a sum of Lorentzian profiles \citep{Handberg2011,Davies2016}. 
The Lorentzian profile is a typical characteristic for solar-like oscillations, as it represents an oscillation mode that experiences damping over time \citep{Anderson1990}. We followed the fitting procedure described in \citet{Liyg2020}.

\subsection{Peakbagging with FAMED}
The peakbagging analysis conducted with the \famed pipeline \citep{Corsaro20} is an automated process based on the use of the public code \diamonds for Bayesian Inference \citep{Corsaro14}\footnote{The public GitHub repository is available at \url{https://github.com/EnricoCorsaro/DIAMONDS}}. The analysis builds on a preliminary step, conducted separately, which is in charge of estimating the level of the background in the stellar PDS. This preliminary phase is performed by means of the public tool \textsc{Diamonds+Background}, which on top of estimating the model free parameters allows to identify an optimal background model, typically consisting of a series of Harvey-like profiles, a flat noise component, and a Gaussian envelope to reproduce the solar-like oscillations \citep{Corsaro17}. The peakbagging procedure instead relies on the exploitation of a multi-modal sampling done with \diamonds, which quickly localizes a large number of relevant frequency peaks in the stellar PDS while ensuring the adoption of a low number of free parameters during the fitting process \citep{Corsaro19}.

\subsection{Peakbagging following Benomar et al. 2013}
The power spectrum analysis is performed using a MCMC algorithm, initially presented in \cite{Benomar2009} and refined to fit pulsations in red giants. The multi-step approach begins by fitting a Gaussian to the mode envelope, with the noise background modelled following the formalism of \cite{Kallinger2014}. The posterior distributions of the noise parameters from this initial fit are then used as priors for the detailed fit of the mode structure. A semi-automated method, relying on the Gaussian mode envelope parameters, provides the initial guesses and priors for the modes parameters, which are assumed to be symmetrical Lorentzians. Two models suitable for RGB analysis are available in the fitting software\footnote{Available at \url{https://github.com/OthmanB/TAMCMC-C}}, and we found that the most constrained model is sufficient to describe the mode pattern observed in the data. This model closely resembles the one used by \cite{Dhanpal22} for training machine learning models. It assumes that frequency variations of the $\ell=0$ and $\ell=2$ modes are described with a cubic spline, as in \citealt{Benomar2013}. However, dipole mixed modes are assumed to strictly follow the asymptotic relation for mixed modes, allowing the extraction of asymptotic parameters such as $q$, $\Delta\nu$ and $\Delta\Pi_1$. 

For Hennes, the model accurately describes the mode patterns, with residuals of the spectrum averaging to $\simeq 1.001$ within the fitting range. Parameters show mild to no multi-modalities and are determined with high precision, as detailed in Table \ref{tab:ColabGlobParams}. For Rogue, however, the posterior distribution for the period spacing is completely uniform over the range [50, 150], indicating that it is not measurable. Consequently, $q$ peaks at a value consistent with zero are within $2\sigma$.

\subsection{Peakbagging with TACO}\label{subapp:TACO}
TACO, the Tools for the Automated Characterisation of Oscillations, is a data driven peakbagging code (\citealt{TACOPrelim}; Hekker et al. in prep). This code aims to perform the full analysis from timeseries to fitted power density spectrum (PDS) with fully identified and characterized oscillations. The code performs a Fourier transform and fits the global features of the PDS using Bayesian methods. On the background corrected PDS, TACO automatically identifies peaks using the Mexican hat wavelet function \citep{andres2018}. These peaks are subsequently fitted using MLE fits. As all significant peaks are fitted for, the Akaike Information Criterion (AIC) is used to identify the most significant peaks that we attribute to oscillation signals. Subsequently, TACO characterizes the fitted peaks broadly using known asymptotic relations. 

The peakbagging results from TACO are available in this remote repository: \url{https://www.erda.au.dk/archives/794d4cf5ad99acf2df21ccd28eca57e1/published-archive.html}.

\subsection{ML analysis of Hennes and Rogue}
We applied a neural network, trained to infer global seismic parameters \dnu, \deltapi, \numax and $q$, to the oscillation spectra of \hennes and \rogue. This convolutional neural network was trained on 5 million synthetic spectra, as described by the asymptotic theory of stellar oscillations. Additional details about the datasets and the neural network can be found in \cite{Dhanpal22} and \cite{Dhanpal23}. Measurements on \hennes and \rogue are described in Table \ref{app:A}.

For evolved red giants like \rogue, $q$ is smaller ($\simeq 0.03$). It has been discussed in \cite{Dhanpal23} that the uncertainty in \deltapi increases at low $q$. This may be due to the decrease in the amplitude of g-dominated mixed modes and the reduction in the transmission factor. Consequently, this star exhibits a large uncertainty of $58$~s.

\begin{table}[h!]
\centering
\caption{Consolidated list of individual frequencies for \hennes. No correction to the Doppler shift has been applied.}
\begin{tabular}{cccc}
\multicolumn{4}{c}{\textbf{Hennes -- \kichennes}} \\ \hline
\multicolumn{1}{c}{\thead{Order \\ $n$}} & 
\multicolumn{1}{c}{\thead{Degree \\ $\ell$}} & 
\multicolumn{1}{c}{\thead{Frequency \\ \makebox[0pt][c]{[$\mu$Hz]}}} & 
\multicolumn{1}{c}{\thead{Uncertainty \\ \makebox[0pt][c]{[$\mu$Hz]}}} \\ \hline
5 & 2 & 68.1678 & 0.2135 \\ 
6 & 0 & 69.5850 & 0.0808 \\ 
6 & 1 & 74.2323 & 0.0757 \\ 
6 & 1 & 74.5773 & 0.0459 \\ 
6 & 2 & 77.2774 & 0.0539 \\ 
7 & 0 & 78.7856 & 0.0628 \\ 
7 & 1 & 82.9593 & 0.0938 \\ 
7 & 1 & 83.5408 & 0.0670 \\ 
7 & 1 & 84.2097 & 0.0974 \\ 
7 & 1 & 85.1689 & 0.0350 \\ 
7 & 2 & 86.9481 & 0.1888 \\ 
8 & 0 & 88.5026 & 0.0946 \\ 
8 & 1 & 92.4132 & 0.1938 \\ 
8 & 1 & 92.9687 & 0.0798 \\ 
8 & 1 & 93.5001 & 0.1104 \\ 
8 & 1 & 93.9274 & 0.0956 \\ 
8 & 1 & 94.5140 & 0.1539 \\ 
8 & 2 & 96.6655 & 0.1482 \\ 
9 & 0 & 98.1491 & 0.1165 \\ 
9 & 1 & 100.8280 & 0.0619 \\ 
9 & 1 & 101.5738 & 0.1190 \\ 
9 & 1 & 102.3327 & 0.0391 \\ 
9 & 1 & 102.8020 & 0.0493 \\ 
9 & 1 & 103.4309 & 0.0602 \\ 
9 & 1 & 104.0100 & 0.0403 \\ 
9 & 1 & 104.6220 & 0.0500 \\ 
9 & 1 & 105.3686 & 0.0709 \\ 
9 & 2 & 106.5278 & 0.1898 \\ 
10 & 0 & 108.0101 & 0.1496 \\ 
10 & 1 & 111.4051 & 0.0215 \\ 
10 & 1 & 112.1443 & 0.0331 \\ 
10 & 1 & 112.8324 & 0.0423 \\ 
10 & 1 & 113.5163 & 0.0414 \\ 
10 & 1 & 114.2211 & 0.0713 \\ 
10 & 1 & 115.0487 & 0.0568 \\ 
10 & 2 & 116.5503 & 0.1358 \\ 
11 & 0 & 118.0876 & 0.1260 \\ 
11 & 1 & 122.3061 & 0.0617 \\ 
11 & 1 & 123.0262 & 0.0528 \\ 
11 & 1 & 123.6331 & 0.0264 \\ 
11 & 2 & 126.4318 & 0.1612 \\ 
12 & 0 & 128.0299 & 0.0696 \\ 
12 & 1 & 133.1081 & 0.0117 \\ 
12 & 1 & 133.9477 & 0.0455 \\ 
\hline
\end{tabular}
\label{tab:FreqAsteroHennes}
\end{table}

\begin{table}[htbp]
\centering
\caption{Consolidated list of individual frequencies for \rogue. No correction to the Doppler shift has been applied.}
\begin{tabular}{cccc} 
\multicolumn{4}{c}{\textbf{Rogue -- \kicrogue}} \\ \hline
\multicolumn{1}{c}{\thead{Order \\ $n$}} & 
\multicolumn{1}{c}{\thead{Degree \\ $\ell$}} & 
\multicolumn{1}{c}{\thead{Frequency \\ \makebox[0pt][c]{[$\mu$Hz]}}} & 
\multicolumn{1}{c}{\thead{Uncertainty \\ \makebox[0pt][c]{[$\mu$Hz]}}} \\ \hline
3 & 2 & 19.3921 & 0.0492 \\ 
4 & 0 & 20.2413 & 0.0304 \\ 
4 & 1 & 22.3579 & 0.0406 \\ 
5 & 2 & 23.4145 & 0.0196 \\ 
5 & 0 & 24.0470 & 0.0146 \\ 
5 & 1 & 26.0382 & 0.0316 \\ 
6 & 2 & 27.2797 & 0.0160 \\ 
6 & 0 & 27.9256 & 0.0102 \\ 
6 & 1 & 30.0742 & 0.0483 \\ 
7 & 2 & 31.3467 & 0.0131 \\ 
7 & 0 & 31.9625 & 0.0096 \\ 
7 & 1 & 34.0882 & 0.0438 \\ 
8 & 2 & 35.3831 & 0.0328 \\ 
8 & 0 & 36.0381 & 0.0143 \\ 
8 & 1 & 38.3250 & 0.0401 \\ 
9 & 2 & 39.6462 & 0.0607 \\ 
9 & 0 & 40.2443 & 0.0536 \\ 
\hline
\end{tabular}
\label{tab:FreqAsteroRogue}
\end{table}

\subsection{Consolidated frequencies for Hennes}
The consolidated list of frequencies based on the considerations in Sect.~\ref{subsec:AsteroHennes} is displayed in Table~\ref{tab:FreqAsteroHennes} for easy access. Note that these frequencies are the output of the peakbagging procedure of \famed and thus the correction to the Doppler shift as proposed by \citet{Davies14} has not been applied.

\subsection{Consolidated frequencies for Rogue}
The consolidated list of frequencies based on the considerations in Sect.~\ref{subsec:AsteroRogue} is displayed in Table~\ref{tab:FreqAsteroRogue} for easy access. Note that these frequencies are the output of the peakbagging procedure described in \ref{subapp:Yaguang} and thus the correction due to the Doppler shift as proposed by \citet{Davies14} has not been applied.

\clearpage

\section{Corner plots of modelling results}\label{app:B}
\begin{figure*}[t]
    \includegraphics[width=17cm]{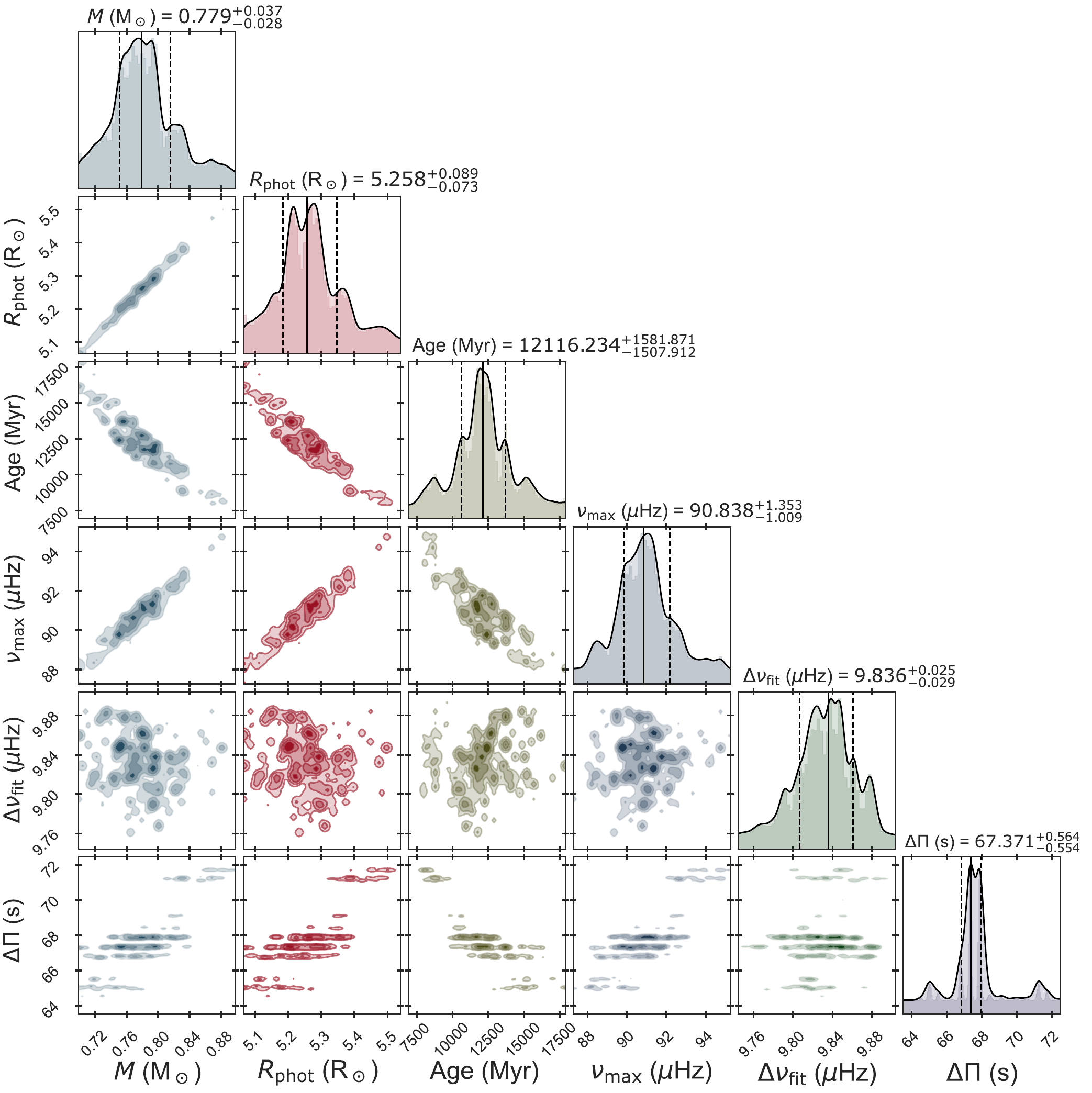}
    \centering
    \caption{Corner plot showing the posterior distributions obtained for the fundamental stellar parameters and global asteroseismic parameters of \hennes.}
    \label{fig:HennesCorner}
\end{figure*}
\begin{figure*}[t]
    \includegraphics[width=17cm]{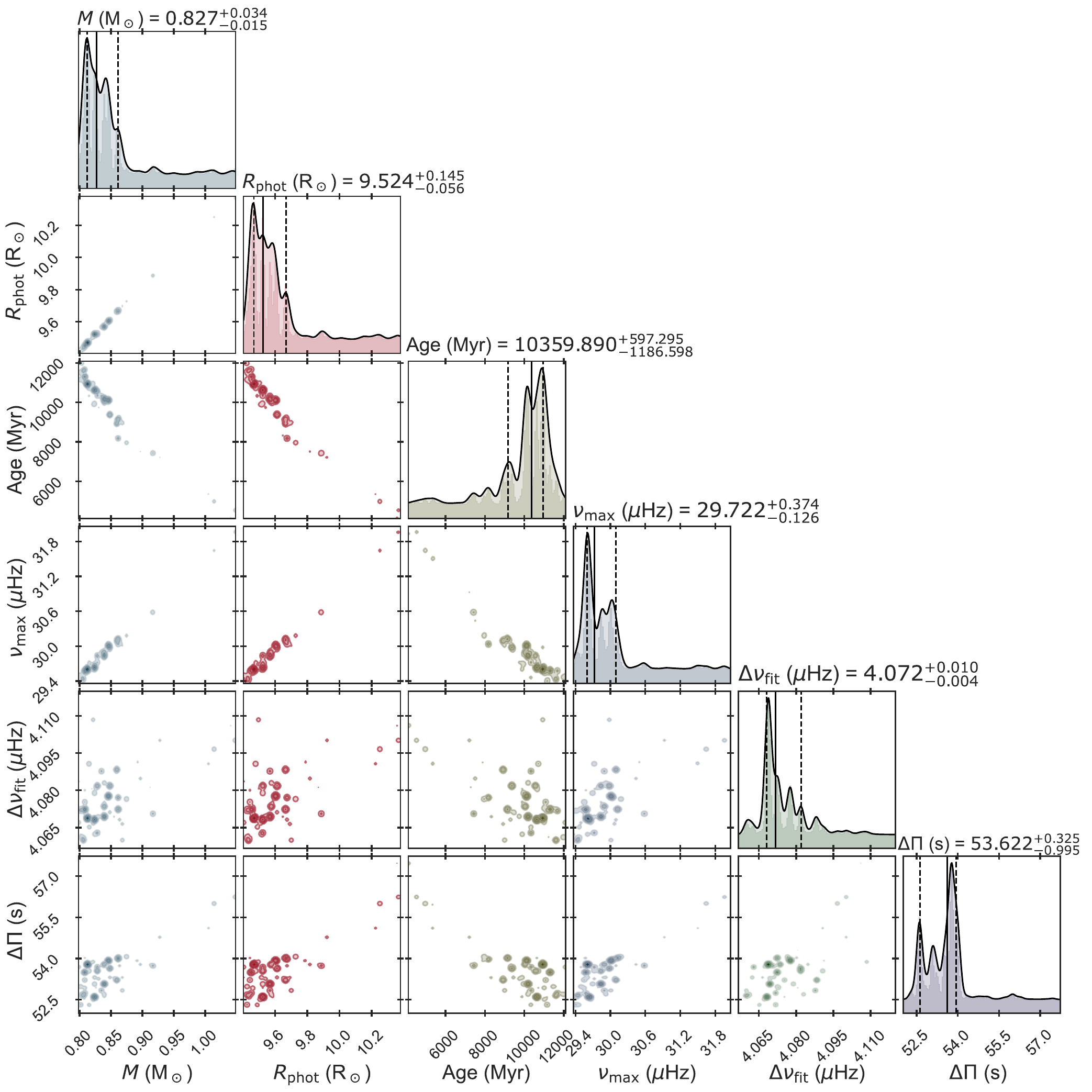}
    \centering
    \caption{Corner plot displaying the posteriors obtained for the fundamental stellar parameters and global asteroseismic parameters from the modelling of \rogue.}
    \label{fig:RogueCorner}
\end{figure*}

This appendix displays the corner plots of the posterior distributions obtained from the grid-based stellar modelling of \hennes and \rogue in Figures \ref{fig:HennesCorner} and \ref{fig:RogueCorner}. The corner plots show each of the stellar parameters displayed in Tables \ref{tab:HennesResults} and \ref{tab:RogueResults} plotted against one another. This allows for visual inspection of correlations between the parameter dimensions. The marginalised posterior distribution of each parameter is showed on the top of their respective column, indicating the median (solid line) and 16th and 84th percentiles (dashed lines), which we will use as central value and uncertainties in this work.

In the modelling we allowed the values of the mixing length parameter $\alpha_\mathrm{mlt}$ and initial helium abundance $Y_\mathrm{ini}$ to vary freely. From the obtained posteriors of Hennes and Rogue, we extracted the obtained values of these parameters. For Hennes, $\alpha_\mathrm{mlt} = 1.716^{+0.155}_{-0.128}$ and $Y_\mathrm{ini}=0.262^{+0.015}_{-0.011}$, while for Rogue $\alpha_\mathrm{mlt} = 1.893^{+0.069}_{-0.050}$ and $Y_\mathrm{ini}=0.251^{+0.005}_{-0.002}$. These values for the mixing length do not strongly deviate from expectations, as they are close to our solar calibrated value of $\alpha_\mathrm{mlt} = 1.786$.  The initial helium estimates are slightly higher than that estimated from a chemical enrichment law $\mathrm{d}Y/\mathrm{d}Z = 1.4$ \citep{Balser06,Brogaard12}, but is consistent within the uncertainties.

\clearpage
\section{Fitting the global asteroseismic parameters}\label{app:C}
\begin{table}[h]
\renewcommand{\arraystretch}{1.3}
\centering
\caption{Modelling results for \hennes when using global asteroseismic parameters. All stellar parameters are given by the median of the posterior distribution, with uncertainties as the 16th and 84th quantiles.}
\begin{tabular}{lccc}
\multicolumn{3}{c}{\textbf{Hennes -- \kichennes}} \\ \hline
\multicolumn{1}{l}{\thead{Stellar parameter}} & \multicolumn{1}{c}{\thead{This work}} & \multicolumn{1}{c}{\thead{Alencastro Puls \\ et al. 2022}} \\ \hline 
$M$ [\solarmass]  &  $0.86^{+0.02}_{-0.02}$      &  $1.01^{+0.02}_{-0.02}$   \\
$R$ [\solarradius]  &  $5.43^{+0.05}_{-0.05}$      &  $5.70^{+0.05}_{-0.05}$   \\
Age [Gyr]        &  $9.2^{+0.8}_{-0.6}$          &  $5.4^{+0.4}_{-0.3}$      \\
\numax [\muHz]     &  $94.08^{+0.73}_{-0.69}$      &  --    \\
\dnu [\muHz]     &  $9.86^{+0.05}_{-0.05}$     &  --    \\
$\Delta\Pi_1$ [s]   &  $69.38^{+0.93}_{-0.75}$      &  --    \\
\end{tabular}
\label{tab:HennesGlobalResults}
\end{table}

\begin{table}
\renewcommand{\arraystretch}{1.3}
\centering
\caption{Modelling results for \rogue when using global asteroseismic parameters. All stellar parameters are given by the median of the posterior distribution, with uncertainties as the 16th and 84th quantiles.}
\begin{tabular}{lcc}
\multicolumn{3}{c}{\textbf{Rogue -- \kicrogue}} \\ \hline
\multicolumn{1}{l}{\thead{Stellar parameter}} & \multicolumn{1}{c}{\thead{This work}} & \multicolumn{1}{c}{\thead{\citet{Pinsonneault24}}} \\ \hline 
$M$ (\solarmass)    &  $1.04^{+0.006}_{-0.003}$      & $1.05\pm 0.040$      \\
$R$ (\solarradius)    &  $10.56^{+0.03}_{-0.03}$      & $10.32\pm 0.18$   \\
Age [Gyr]          &  $4.81^{+0.05}_{-0.14}$      & $4.85^{+0.60}_{-0.52}$   \\
\numax [\muHz]     &  $30.83^{+0.09}_{-0.05}$    & --       \\
\dnu [\muHz]     &  $4.02^{+0.02}_{-0.01}$     & --       \\
$\Delta\Pi_1$ [s]   &  $56.11^{+0.32}_{-0.39}$    & --       \\
\end{tabular}
\label{tab:RogueGlobalResults}
\end{table}

\begin{figure}[t]
    \resizebox{\hsize}{!}{\includegraphics[width=\linewidth]{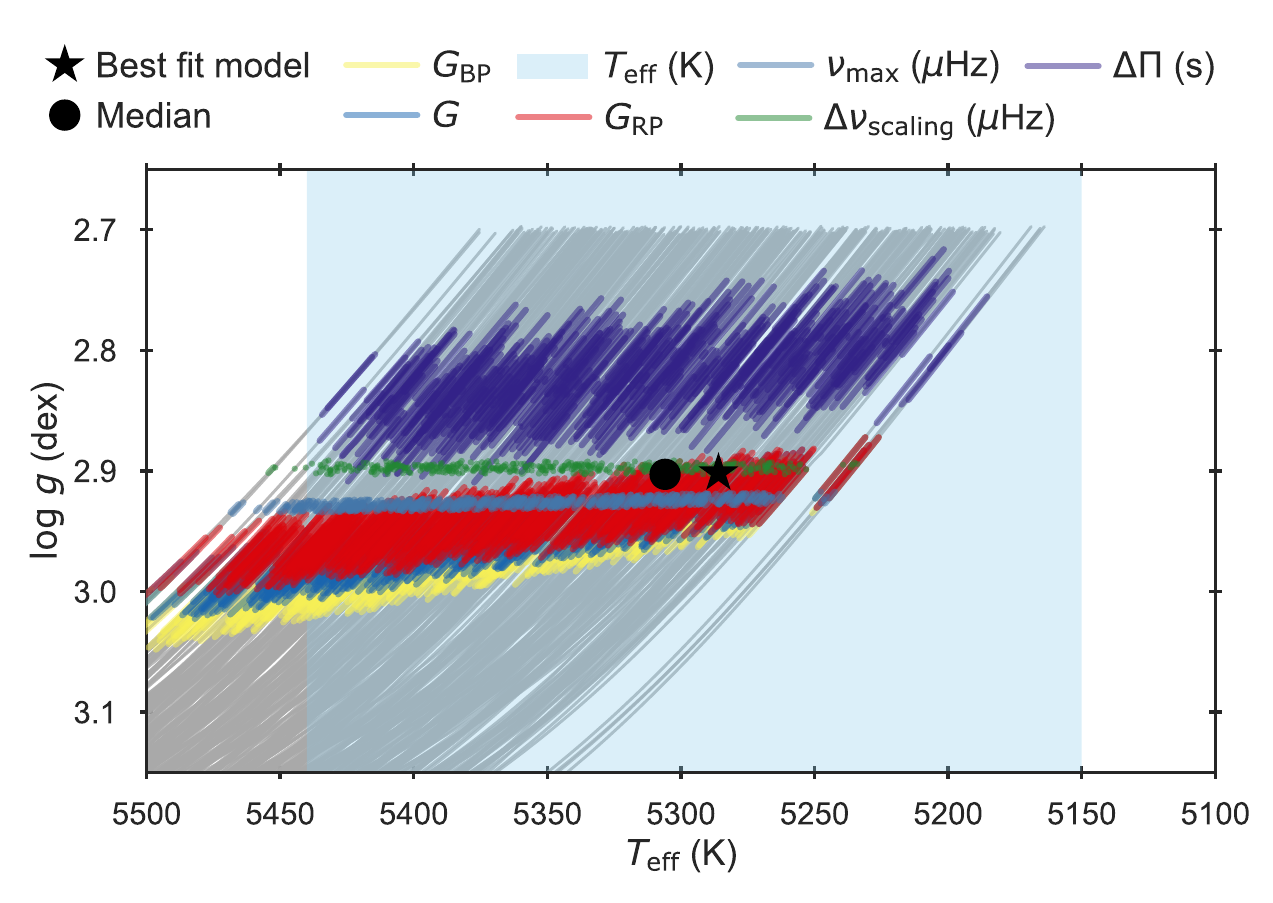}}
    \centering
    \caption{Kiel diagram of \hennes for a fit to the global asteroseismic parameters, displaying a representative number of stellar tracks in the grid. The observed constraints on the fitted parameters are overlaid to show a clear lack of overlap with the median and BFM from the posterior.}
    \label{fig:HennesKielGlobal}
\end{figure}
\begin{figure}[t]
    \resizebox{\hsize}{!}{\includegraphics[width=\linewidth]{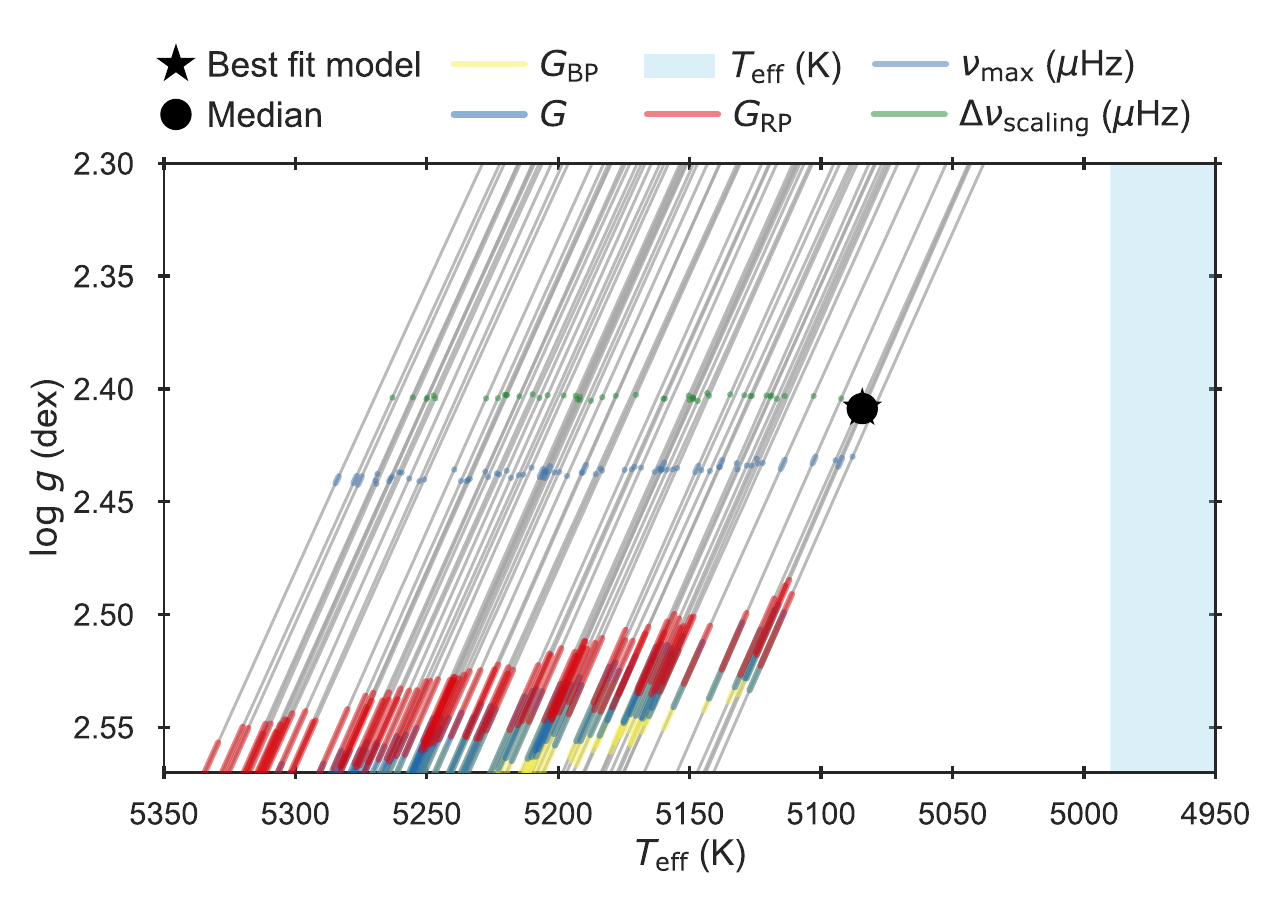}}
    \centering
    \caption{Kiel diagram of \rogue for a fit to the global asteroseismic parameters, displaying a representative number of stellar tracks in the grid. The observed constraints on the fitted parameters are overlaid to show a clear lack of overlap with the median and BFM from the posterior.}
    \label{fig:RogueKielGlobal}
\end{figure}

In order to emphasise the contrast to the previously best-possible case when asteroseismically modelling giant stars, we here present the results from a fit to the global asteroseismic parameters \dnu, \numax and $\Delta\Pi_1$. The same classical parameters as in Sect.~\ref{sec:Results} are included in the fit. For \rogue, due to the reasons described in Sect.~\ref{subsec:AsteroRogue}, we do not attempt to fit $\Delta\Pi_1$.

Figures~\ref{fig:HennesKielGlobal} and \ref{fig:RogueKielGlobal} show the Kiel diagrams for \hennes and \rogue, respectively. A clear lack of overlap between the observed constraint bands are seen, i.e. a tension between the observables when translated to model space. This was the same situation that the prior efforts of V. Silva Aguirre et al. (unpublished) initially faced when performing the modelling attempts ten years ago. Tables \ref{tab:HennesGlobalResults} and \ref{tab:RogueGlobalResults} show the stellar parameters obtained for the modelling. Both \hennes and \rogue are recovered as vastly different stars from the ones found by modelling them utilising individual frequencies (see Sect~\ref{sec:Results}). Note that the lower uncertainties recovered here, when compared to the ones found by the individual frequency modelling, occur due to the low uncertainties on the global asteroseismic parameters and the lacking overlap between them in model space.

A final interesting mention is that when we fit the global asteroseismic parameters but omit \numax, we can generally recover an overlap between the fitted constraints. This provides posteriors much more similar to the ones from individual frequency modelling. The posteriors are wider, which results in larger errors on the determined parameters, but encompasses the solution found by the individual frequency modelling. This further speaks to the unreliability of the \numax scaling relation for low-metallicity evolved stars (see Sect.~\ref{subsec:numaxDiscrep}).

\clearpage
\section{Orbital Solutions for Hennes and Rogue}\label{app:D}
Table~\ref{tab:orbitalparams} lists the orbital properties for \hennes and \rogue as described in Sect.~\ref{sec:Discussion}. \emph{GC} refers to Galactocentric quantities, while \emph{HC} are heliocentric.

\renewcommand{\arraystretch}{1.4}
\begin{table}
\centering
\caption{Summarised orbital parameters for Hennes and Rogue.}
\label{tab:orbitalparams}
\begin{tabular}{lll}
\multicolumn{1}{l}{\thead{Orbital parameter}} & \multicolumn{1}{c}{\thead{Hennes \\ \kichennes}} & \multicolumn{1}{c}{\thead{Rogue \\ \kicrogue}} \\ \hline
$X$ (GC) (kpc) & $7.85_{-0.01}^{+0.01}$   &   $7.59_{-0.01}^{+0.02}$   \\
$Y$ (GC) (kpc) & $1.54_{-0.03}^{+0.02}$   &   $2.13_{-0.06}^{+0.05}$   \\
$Z$ (GC) (kpc) & $0.3_{-0.0}^{+0.0}$   &   $0.33_{-0.01}^{+0.01}$   \\
$U$ (HC) (km/s) & $-26.0_{-0.4}^{+0.4}$   &   $-87.4_{-1.0}^{+1.3}$   \\
$V$ (HC) (km/s) & $0.4_{-0.2}^{+0.2}$   &   $-191.6_{-1.3}^{+1.9}$   \\
$W$ (HC) (km/s) & $-8.9_{-0.1}^{+0.3}$   &   $118.4_{-4.2}^{+3.0}$   \\
$R$ (GC) (kpc) & $7.999_{-0.001}^{+0.001}$   &   $7.883_{-0.0}^{+0.001}$   \\
$\phi$ (GC) (kpc) & $2.95_{-0.0}^{+0.0}$   &   $2.87_{-0.01}^{+0.01}$   \\
$Z$ (GC) (kpc) & $0.3_{-0.0}^{+0.0}$   &   $0.33_{-0.01}^{+0.01}$   \\
$v_{R}$ (GC) (km/s) & $59.6_{-1.3}^{+1.0}$   &   $84.5_{-1.1}^{+0.7}$   \\
$v_{\phi}$ (GC) (km/s) & $226.3_{-0.3}^{+0.4}$   &   $19.6_{-2.2}^{+2.9}$   \\
$v_z$ (GC) (km/s) & $-1.6_{-0.1}^{+0.3}$   &   $125.9_{-4.2}^{+3.0}$   \\
$J_R$ (km/s kpc) & $51_{-2}^{+1}$   &   $503_{-3}^{+9}$   \\
$L_z$ (km/s kpc) & $1864.7_{-2.6}^{+3.8}$   &   $159.0_{-17.7}^{+23.7}$   \\
$J_z$ (km/s kpc) & $3.0_{-0.1}^{+0.1}$   &   $352.2_{-32.4}^{+28.4}$   \\
$e$ (--) & $0.194_{-0.004}^{+0.003}$   &   $0.909_{-0.013}^{+0.009}$   \\
$z_{\textup{max}}$ (kpc) & $0.38_{0.001}^{0.001}$   &   $6.67_{-0.05}^{+0.04}$   \\
$r_{\rm peri}$ (kpc) & $7.057_{0.004}^{0.003}$   &   $0.445_{-0.005}^{+0.008}$   \\
$r_{\rm ap}$ (kpc) & $10.45_{0.008}^{0.002}$   &   $9.366_{-0.01}^{+0.008}$   \\
Energy ((km/s)$^2$) & $-157510_{60}^{61}$   &   $-173911_{-669}^{+373}$   \\
$L_{\rm total}$ (km/s kpc) & $1811_{2}^{4}$   &   $980_{-35}^{+18}$   \\
\end{tabular}
\end{table}

\section{Frequency fitting algorithm of \basta}\label{app:E}
In \basta \citep{Aguirre22}, the pipeline assesses the likelihood of a stellar model given the observational constraint of a set of individual mode frequencies in multiple steps.

In order to save computation time, \basta can utilise the fact that the mode frequencies of the model and the observed mode frequencies need to be rather close for the likelihood to be significant. The user specifies a distance in frequency space, parametrised as a fraction of the observed large frequency separation. This distance is transformed into an asymmetric prior for which models only pass if the $\ell=0$ mode with the same radial order as the lowest observed radial mode is close to the observed counterpart. The asymmetric nature of this prior describes the expectation that due to the asteroseismic surface effect the model frequencies have greater values than the observed ones \citep{Jørgen88}.

As it is not guaranteed that all modes present in the star have observable visibilities, it is possible for a good model to have more modes than the set of observed frequencies, but not in reverse: a model that accurately describes the structure of the star cannot physically have a pattern of mode frequencies with less modes than what is observed, given that we believe every mode frequency in the set of observations to be real. Based on this principle, \basta matches the observed modes to model counterparts using an algorithm that does not depend on the assigned mode identification of radial modes, but instead on the observed pattern. This is described in more detail in \citet{Aguirre22} or alternatively in Sect.~3.1 of \citet{Ball20}, but briefly summarised \basta matches the radial frequencies assuming that the radial order of the lowest radial mode is correct. Using the radial frequencies as boundaries for different segments and thus avoiding the issue that the surface effect can impact on matching in pure frequency space, it then matches the observed frequencies to model counterparts using the inverse mode inertia as a proxy for visibility \citep{Benomar14}. 

A single modification was made in this procedure for the results presented in this work, as we consider evolved RGB stars. The offset between the boundaries of the different sets is only applied if a single observed non-radial frequency is available, e.g. as is the case for MS stars. However, as this work considers evolved RGB stars where all dipole modes are mixed modes with significant g-mode components, this frequency shift will result in unrealistic displacements of the $\ell=1$ modes during the matching. We have therefore forced the shift to be zero when matching the dipole modes, which is only enforced for Rogue, as we consider multiple dipole modes per acoustic mode order for Hennes. We note that these changes will impact the exact best-fitting model but not significantly the overall posterior distribution.

\end{appendix}

\end{document}